\input harvmac

    \def\b{\beta}              
\def\d{\delta}
    \def\e{\varepsilon}        

\def\g{\gamma}               

   \def\m{\mu}         \def\n{\nu}        
\def\r{\rho}
             
\def\p{\psi}
      \def\s{\sigma}      \def\S{\Sigma}     
\def\th{\theta}
\def\t{\tau}

\def\CD{{\cal D}}

\def\CG{{\cal G}}

\def\CN{{\cal N}}

\def\CV{{\cal V}}
\def\CW{{\cal W}}
\def\CX{{\cal X}}
\font\teneufm=eufm10
\font\seveneufm=eufm7
\font\fiveeufm=eufm5
\newfam\eufmfam
\textfont\eufmfam=\teneufm
\scriptfont\eufmfam=\seveneufm
\scriptscriptfont\eufmfam=\fiveeufm

\font\teneusm=eusm10
\font\seveneusm=eusm7
\font\fiveeusm=eusm5
\newfam\eusmfam
\textfont\eusmfam=\teneusm
\scriptfont\eusmfam=\seveneusm
\scriptscriptfont\eusmfam=\fiveeusm

\font\tenmsx=msam10
\font\sevenmsx=msam7
\font\fivemsx=msam5
\font\tenmsy=msbm10
\font\sevenmsy=msbm7
\font\fivemsy=msbm5
\newfam\msafam
\newfam\msbfam
\textfont\msafam=\tenmsx  \scriptfont\msafam=\sevenmsx
  \scriptscriptfont\msafam=\fivemsx
\textfont\msbfam=\tenmsy  \scriptfont\msbfam=\sevenmsy
  \scriptscriptfont\msbfam=\fivemsy

\def\msbm#1{{\fam\msbfam\relax#1}}
\font\tenbifull=cmmib10 
\font\tenbimed=cmmib10 scaled 800
\font\tenbismall=cmmib10 scaled 666
\textfont9=\tenbifull \scriptfont9=\tenbimed
\scriptscriptfont9=\tenbismall

\def\Bv{{\fam=9{\mathchar"7120 } }}

\def\cmp#1#2#3{Comm.\ Math.\ Phys.\ {{\bf #1}} {(#2)} {#3}}
\def\pl#1#2#3{Phys.\ Lett.\ {{\bf #1}} {(#2)} {#3}}
\def\npb#1#2#3{Nucl.\ Phys.\ {{\bf #1}} {(#2)} {#3}}
\def\prd#1#2#3{Phys.\ Rev.\ {{\bf #1}} {(#2)} {#3}}

\def\ijmp#1#2#3{Int.\ J.\ Mod.\ Phys.\ {{\bf #1}} 
{(#2)} {#3}}

\def\im#1#2#3{Invent.\ Math.\ {{\bf #1}} {(#2)} {#3}}
\def\plms#1#2#3{Proc.\ London Math.\ Soc.\ {{\bf #1}} 
{(#2)} {#3}}

\def\jgp#1#2#3{J.\ Geom.\ Phys.\ {{\bf #1}} {(#2)} {#3}}

\def\rd{\partial}

\def\darr#1{\raise1.5ex\hbox{$\leftrightarrow$}
\mkern-16.5mu #1}

\def\fr#1#2{{\textstyle{#1\over#2}}}
\def\Fr#1#2{{#1\over#2}}
\def\tr{\hbox{Tr}\,}

\def\roughly#1{\raise.3ex\hbox{$#1$\kern-.75em
\lower1ex\hbox{$\sim$}}}

\def\pr{\prime}

\def\ack{\bigbreak\bigskip\bigskip\centerline{
{\bf Acknowledgments}}\nobreak}
\baselineskip=16pt plus 1.2pt minus .6pt
\newskip\normalparskip
\normalparskip = 4pt plus 1.0pt minus .5pt
\parskip = \normalparskip
\parindent= 16pt

\font\Titlerm=cmr10 scaled\magstep3
\lref\Monads{
J.-S.~Park,
{\it Monads and D-instantons},
\npb{B 494}{1997}{198}
hep-th/9612096.
}
\lref\ADHM{
G.~Horrocks,
{\it Vector bundles on the punctured spectrum of 
a local ring},
\plms{14}{1964}{684}
\semi
W.~Barth,
{\it Some properties of stable rank-2 vector bundles on 
$P_n$},
\im{42}{1977}{63}
\semi
M.~Atiyah, V.~Drinfeld, N.~Hitchin and Yu.~Manin,
{\it Construction of instantons},
\pl{65A}{1978}{185}
\semi
S.K.~Donaldson, 
{\it Instantons and geometrical invariant theory},
\cmp{93}{453}{1984}
}
\lref\Nahm{
W.~Nahm,
{\it Multimonopoles in the ADHM construction,}
in ``Proc.~Symp. on particle physics'',
eds. Z.~Horvath et al., (Vsiegrad, 1981);
{\it Construction of all self-dual monopoles by the ADHM method},
in ``Monopoles in quantum field theory''.
eds. N.~Craigie et al., (World Scientific, Singapore, 1982);
{\it Self-dual monopoles and calorons,}
in Lecture Notes in Physics, vol.~201,
eds.~G.~Denardo et al., (Springer 1984)
}

\lref\VW{
C.~Vafa and E.~Witten,
{\it A strong coupling test of $S$-duality},
\npb{B 431}{1994}{3}, hep-th/9408074
}
\lref\DM{
R. Dijkgraaf, G. Moore,
{\it Balanced topological field theories},
hep-th/9608169 
}
\lref\DPS{
R.~Dijkgraaf, J.-S.~Park and B.J.~Schroers,
{\it Balanced topological Yang-Mills theory},
to appear.
}
\lref\Park{
J.-S.~Park,
{\it N=2 topological Yang-Mills theory
on compact K\"{a}hler surfaces},
\cmp{163}{1994}{113};
{\it Holomorphic Yang-Mills theory on compact
K\"{a}hler manifolds};
\npb{B 423}{1994}{559}
\semi
S.~Hyun and J.-S.~Park,
{\it N=2 topological Yang-Mills theories
and Donaldson's polynomials},
J.~Geom.~Phys.~ to appear, hep-th/9404009
}
\lref\Pol{
J.~Polchinski,
{\it Dirichlet-branes and Ramond-Ramond charges},
hep-th/9510017 
}
\lref\BFSS{
T. Banks, W. Fischler, S.H. Shenker, and L. Susskind,
{\it M theory as a matrix model: a conjecture},
Phys.~Rev.~{\bf D 55} (1997) 5112, hep-th/9610043.
}
\lref\Taylor{
W.~Taylor,
{\it  D-brane field theory on compact spaces},
\pl{B 394}{1997}{283},  hep-th/9611042 
}
\lref\DMVV{
R.~Dijkgraaf, E.~ Verlinde and H.~Verlinde,
{\it BPS spectrum of the five-brane and black-hole
entroy},
hep-th/9607026
\semi
R.~Dijkgraaf, G.~Moore, E.~ Verlinde and H.~Verlinde,
{\it Elliptic genera of symmetric products and second
quantized strings},
hep-th/9608096
}
\lref\Motl{
L. Motl,
{\it Proposals on nonperturbative superstring interactions},
hep-th/9701025.
}
\lref\BSs{
T. Banks and N. Seiberg,
{\it Strings from matrices},
hep-th/9702187.
}
\lref\BSb{
T. Banks and N. Seiberg,
{\it Branes from matrices},
\npb{B 490}{1997}{91},
hep-th/9612157
}
\lref\DVV{
R. Dijkraaf, E. Verlinde, and H. Verlinde,
{\it Matrix string theory},
hep-th/9703030;
R. Dijkraaf, E. Verlinde, and H. Verlinde,
{\it 5D black holes and matrix strings},
hep-th/9704018.
}
\lref\Mhet{
T.~Banks and L.~Motl,
{\it Heterotic strings from matrices},
hep-th/9703218
\semi
D.A.~Lowe,
{\it Heterotic matrix string theory},
hep-th/9704041
\semi
S.-J.~Rey,
{\it Heterotic matrix strings and their
interactions},
hep-th/9704158
\semi
P.~Horava,
{\it Matrix theory and heterotic strings on tori},
hep-th/9705055
\semi 
N.~Kim and S.-J.~Rey,
{\it M(atrix) theory on $T^5/Z_2$ orbifold and
five-branes},
hep-th/9705132
}
\lref\WittenE{
E.~Witten,
{\it Bound states of strings and $p$-branes},
\npb{B 460}{1996}{335}, hep-th/9510135.
}
\lref\Witten{ 
E.~Witten,
{\it Topological quantum field theory},
\cmp{117}{1988}{353};
{\it Introduction to cohomological field theories},
\ijmp{A6}{1991}{2273};
{\it Topological Sigma Models},
\cmp{118}{1988}{411}.
}
\lref\WittenG{
E.~Witten,
{\it Two dimensional gauge theories revisited},
\jgp{9}{1992}{303}, hep-th/9204083.
}
\lref\WittenV{
E.~ Witten,
{\it String theory dynamics in various dimensions},
\npb{B 443}{1995}{85},
hep-th/9503124.  
}
\lref\SWb{
N.~Seiberg and E.~Witten,
{\it Monopoles, duality, and chiral symmetry breaking in 
$N=2$ supersymmetric QCD},
\npb{B 431}{1994}{484}, hep-th/9408099.
}

\lref\HT{
C.~ Hull and P.~ K.~ Townsend, 
{\it Unity of superstring dualities},
\npb{B 438}{1995}{109}, hep-th/9410167  
}
\lref\GSW{
M.~Green, J.H.~Schwarz and E.~Witten,
{\it Superstring theory}, vol I,II,
Cambridge Univ.~Press 1987.
}
\lref\GSO{
F.~Gliozzi, J.~Sherk and D.~Olive,
{\it Supersymmetry, supergravity theories and the dual spinor
model},
\npb{B 112}{1977}{253}
}
\lref\RNS{
P.~Ramond,
{\it Dual theory for free fermions},
\prd{D 3}{1971}{2415}\semi
A.~Neveu and J.H.~Schwarz,
{\it Factorizable dual model of pions},
\npb{B 31}{1971}{86}.
}
\lref\GS{
M.~Green and J.H.~Schwarz,
{\it Supersymmetric dual string theory},
\npb{B 181}{1981}{502};
{\it Supersymmetrical string theories},
\pl{B 109}{1982}{444}.
}
\lref\SCH{J.H.~Schwarz,
{\it The power of M theory},
\pl{B 367}{1996}{97},
hep-th/9510086
}
\lref\AS{
P.S.~Aspinwall,
{\it Some relationships between dualities in string theory},
Nucl. Phys. Proc. Suppl. {\bf 46} (1996) 30, hep-th/9508154 
\semi
J.H.~Schwarz
{\it  An $SL(2,Z)$ multiplet of type IIB superstrings},
hep-th/9508143.
}
\lref\BRS{
M.~Rozali, {\it  Matrix theory and U-duality in seven dimensions},
hep-th/9702136
\semi
M.~Berkooz, M.~Rozali, and N.~Seiberg,
{\it  Matrix description of M-theory on $T^4$ and $T^5$,
}  hep-th/9704089
\semi
N.~Seiberg,
{\it Matrix description of M-theory on $T^5$ and $T^5/Z_2$},
hep-th/9705221.
}
\lref\HW{
P.~Horava and E.~Witten,
{\it 
 Heterotic and type I string dynamics from eleven dimensions},
\npb{B 460}{1996}{506},  hep-th/9510209.
}

\font\Titlerm=cmr10 scaled\magstep3
\nopagenumbers
\rightline{THU-97/16, UVA-ITFA/97/20}
\rightline{hep-th/9706130
}

\vskip .3in

\centerline{\fam0\Titlerm 
Monads,  Strings, and M Theory
}
\vskip 0.3in

\tenpoint\vskip .1in\pageno=0

\centerline{
Christiaan 
Hofman\footnote{$^{\dagger}$}{e-mail: c.m.hofman@fys.ruu.nl}}
\vskip .1in
\centerline{\it
Institute for Theoretical Physics}
\centerline{\it
University of Utrecht}
\centerline{\it
Princetonplein 5, 3508 TA Utrecht}
\vskip .2in
\centerline{ 
Jae-Suk Park\footnote{$^{\dagger\dagger}$}{e-mail: park@phys.uva.nl}}
\vskip .1in
\centerline{\it Institute for Theoretical Physics}
\centerline{\it University of Amsterdam}
\centerline{\it Valckenierstraat 65, 1018 XE Amsterdam}

\vskip .2in

\noindent
\abstractfont

The recent developments in string theory suggest that
the space-time coordinates should be generalized to non-commuting
matrices. 
Postulating this suggestion as the fundamental
geometrical principle, we formulate a candidate for covariant second
quantized RNS superstrings as a topological
field theory in two dimensions. Our construction is
a natural non-Abelian extension of the RNS string.
It also naturally leads to a model with manifest $11$-dimensional
covariance, which we conjecture to be a formulation
of M theory. The non-commuting
space-time coordinates of the strings are further generalized to 
non-commuting anti-symmetric tensors.
The usual space-time picture and the free superstrings
appear only in certain special phases of the model.
We derive a simple set of algebraic equations, which
determine the moduli space of our model.
We test some aspects of our conjectual M theory 
for the case of compactification
on $T^2$.

\Date{June, 1997}

\baselineskip=14pt plus 1.2pt minus .6pt

\newsec{Introduction}

It is of no doubt that 
superstring theory has an underlying higher symmetry unifying
those of general relativity and Yang-Mills theory \GSW.
Furthermore, the current understanding
of a web of dualities \HT\WittenV\ imply that
all known superstring theories
including $11$-dimensional supergravity should be different
weak-coupling limits of one underlying quantum theory of some 
Mystery
kind \WittenV\SCH.
Unfortunately, the precise nature of
such an underlying theory still remains obscure. 
A strong hint pointing towards an underlying geometrical principle 
of superstrings has emerged from
the dramatic revival of D-branes by Polchinski \Pol.
The description of D-branes, as originally pointed
out by Witten, suggests that {\it the spacetime coordinates
of strings should be treated as non-commuting matrices} \WittenE.
This consideration eventually
led to the program of matrix theory originated from the 
proposal of 
matrix M theory by Banks-Fischler-Shenker-Susskind \BFSS. 
More recently
Motl \Motl, Banks-Seiberg \BSs, 
and Dijkraaf-Verlinde-Verlinde \DVV\ 
developed matrix string theory by compactifying
M(atrix) theory on a circle \BFSS\Taylor. 
Those  formulations may be viewed as non-perturbative
second quantized Green-Schwarz strings \GS\ in the light-cone
gauge.

The purpose of this paper is to initiate a program
toward M-theory
closely related with the manifestly 
covariant Ramond-Neveu-Schwarz (RNS) formulation
of string \RNS. We believe that the non-commutative nature
of spacetime coordinates of strings is clearly directing
us to formulate superstring theory in {\it
a phase in which general covariance, as well as other
higher symmetry, is unbroken}.  The latter proposal
was made by Witten almost a decade ago 
after introducing a new type of generally
covariant quantum field theory called topological
field theory (TFT) \Witten.
In fact, there are
many similarities between the RNS string and 
TFT. From the spacetime viewpoint, the world-sheet
super-charges transform as scalars, which property is a hall-mark
of TFT. It is one of string magics that the
RNS formulation of string leads to space-time supersymmetry
after the GSO projection \GSO.
It is very natural to relate the non-commutative nature
of ``space-time'' coordinates with the strings 
in the  unbroken phase of higher symmetry. 
The purpose of this paper is to demonstrate that
superstring theory can indeed be formulated starting from the 
above 
two suggestions. 
Furthermore, our construction will naturally lead us to
an underlying model with manifest $11$-dimensional
covariance. Here the non-commutative ``space-time
coordinates'' of strings will be further generalized
to non-commutative anti-symmetric tensors. 
The usual space-time picture and the free superstrings
appear in the various limits of the model after
compactifications.

In Sect.~$4$,
we start from a system of ten $N\times N$ matrix functions 
$X^\m(\s^+,\s^-)$ which are functions of two parameters $\s^\pm$
parameterizing a cylinder $\S=\msbm{S}^1\times \msbm{R}$
with trivial canonical line bundle. 
They carry global $SO(9,1)$ vector indices $\m=0,\ldots,9$.
We endow our system with the natural metric
$$
\left|\d X\right|^2 
= \int d\s^+ d\s^-g_{\m\n}
\tr\left(\d X^{\m} \d X^\n\right),
$$
where $g_{\m\n}$ is the Minkowski metric
with signature $(9,1)$. Following the general idea
of topological field theory (TFT) \Witten,
we will construct an almost unique theory by gauge fixing 
the ``world-sheet'' and ``spacetime'' Poincar\'e  symmetries.
In particular, the obvious symmetry for arbitrary shifts
$X^\m \rightarrow X^\m +\d X^\m$ in the ``spacetime'' viewpoint
implies that we are dealing with a topological field theory
on the ``world-sheet''.
Now the most natural object to study in
our system is the equivariant cohomology.
It turns out that the most suitable tool
is {\it the balanced equivariant cohomology} formalized
by Dijkraaf and Moore \DM.
This is an extremely powerful and simple tool which leads
to an almost unique construction of corresponding TFT
called Balanced TFT (BTFT). A typical example of a BTFT
is the twisted $\CN=4$ super-Yang-Mills theory
studied by Vafa and Witten \VW. 
Our equivariant cohomology can be summarized
by a transformation law $Q_\pm X^\m =i\p^\m_\pm$
and the following commutation relations
between the two generators $Q_\pm$
$$
Q_+^2 = -i\Fr{\rd}{\rd\s^+}-i\d_{\phi_{++}},\qquad
\{Q_+,Q_-\} = -2i\d_{\phi_{+-}},\qquad
Q_-^2 = -i\Fr{\rd}{\rd\s^-}-i\d_{\phi_{--}},
$$
where $\d_{\phi}$ denote the $U(N)$ transformation
generated by $\phi$.
One can regard $Q_\pm$ as the BRST-like charges
for the symmetry of the arbitrary shift of $X^\m$
which are nilpotent modulo a $U(N)$ gauge transformation and 
translations along $\s^\pm$. The ``world-sheet'' Lorentz
invariance will be realized by global ghost number symmetry,
which should be anomaly-free.
We have a unique realization of the algebra
and the action functional satisfying our criterion.
We will claim that the resulting theory
describes a covariant second quantized  
Ramond-Neveu-Schwarz (RNS) string in the unbroken phase.

Our model has a free string limit where the original RNS
string is recovered. 
The equivariant cohomology generators
$Q_\pm$ will be the left and right world-sheet super charges.
The ghost fields $\p^\m_\pm$ for the shift 
$X^\m\rightarrow X^\m +\d X^\m$ will be the left and right moving
world-sheet fermions. The direct relation of the  RNS formalism 
rather
than the space-time supersymmetric Green-Schwarz (GS) formulation
is not surprising.
In fact, our formulation is a natural and presumably
unique generalization
of the RNS superstring to incorporate the 
non-commutative ``spacetime'' coordinates of strings.
We will argue that the transition between
the unbroken and broken phases of general covariance
should be  explained by some of the standard quantum 
properties of RNS superstring. 

Our construction will inevitably lead us to introduce an 
anti-symmetric
tensor of rank $2$. 
We will argue that it is required and compatible with the
existence of off-diagonal parts of ``space-time coordinates''.
Our construction will naturally lead  to an underlying theory
with {\it manifest eleven-dimensional covariance}, discussed
in Sect.~$5$. 
The theory is again a stringy BTFT but with
anti-symmetric tensors as   ``space-time coordinates'' of strings.
We will show that the free RNS string appear in a
limit after compactifying the model on a circle.
By compactifying further on a circle, we will show
the emergency to two types of string limits. 
The $S$-duality of type IIB string will be manifest
in our formulations. The anti-symmetric tensor
``coordinates of strings'' $B^{IJ}(\s,\t)$ is somewhat
analogous to the membrane in M theory. This motivates
us to introduce a new rank $5$ anti-symmetric
tensor $J^{IJKLM}(\s,\t)$ as the  five brane in M theory.
We again define a unique extension with the new
degree of freedom. This will lead us to find
the most important equations in our paper,
$$
\eqalign{
[B_{IK},B_J{}^K] +\b
[J_{IKLMN},J_{J}{}^{KLMN}]=0,\cr
[B^{IJ}, J_{J}{}^{KLMN}]=0.\cr
}
$$
Our conjecture  will be that the moduli space of M theory 
is described by the above equations.

In Sect.~$2$ we discuss
the case of constant matrices as the warm-up example,
which has some interests in its own right. We will review
some relevant properties of the balanced equivariant
cohomology and construct, presumably, the simplest 
balanced topological field theory. 
In Sect.~3,
we will also consider four-dimensional settings of our 
constructions. We will discuss some close relations with
the balanced topological Yang-Mills theory, BTYMT in short, 
(the Vafa-Witten
model of twisted $\CN=4$ SYM theory) in four-dimensions.
We will argue that BTYMT describes a certain sub-sector
of four-dimensional strings in the unbroken phase of
general covariance. Here the anti-symmetric tensor fields
will play an important role when relating with the monads 
(the ADHM)
construction of instantons. 
We will use some crucial results of DVV \DVV\ for
interpreting our model as a  second quantized superstring
theory. In our viewpoint, they also demonstrated how
some of the known properties of strings can be seen to arise in the
unbroken phase.

\newsec{Almost Universal Monads}

Throughout this paper we will consider a system (or space) 
$\overline{W}$ of ten  
 matrices $X^\m$ where  
$\m,\n = 0,\ldots, 9$, in the adjoint of an $U(N)$ group \foot{In general we will allow
a matrix $X^\m$ to {\it degenerate}.
This is analogous to the extension of vector bundles
to sheaves.}.
There is a natural $U(N)$ symmetry on acting in this space
\eqn\haa{
X^\m \rightarrow g X^\m g^{-1},\qquad g\in U(N). 
}
We postulate a $SO(9,1)$ global symmetry acting on the index 
$\m,\n=0,\ldots,9$.
Under $SO(9,1)$ the $X^\m$ transform as components of a vector.
On $\overline W$ there is a natural metric which is 
invariant under $U(N)\times SO(9,1)$
\eqn\hab{
\left|\d X\right|^2 =\tr\left(\eta_{\m\n} \d X^\m \d X^\n\right),
}
where $\eta_{\m\n}$ denotes the usual Minkowski metric
with signature $(9,1)$. 

We want to construct
a theory with ``spacetime'' Poincar\'e invariance as
well as $U(N)$ symmetry.
For the $U(N)$ symmetry, we demand the system $\{X^\m\}$
to be equivalent to the system $\{X^{\pr\m}\}$ if
they are related by $X^{\pr\m} = g X^\m g^{-1}$, for $g\in U(N)$. 
In general, we can always associate a center of mass coordinate to 
the $X^\m$ in
$\msbm{R}^{9,1}$ by $x^\m = N^{-1}\tr X^\m$. The translations
of the base spacetime $\msbm{R}^{9,1}$ act on the matrices
$X^\m$ by $X^\m\rightarrow X^\m + w^\m\msbm{I}_N$.
Together with the global $SO(9,1)$ symmetry, we interpret
the above as the ``spacetime'' Poincar\'e symmetry.
The actual spacetime picture
emerges when all of the $X^\m$ commute with each other,
 hence can be
simultaneously diagonalized as $X^\m = diag(x^\m_\ell)$.
By regarding the eigenvalues $x^\m_\ell$ as coordinates of
points (instantons) $x_\ell$ in $\msbm{R}^{9,1}$ we get
indistinguishable $N$-tuple of points in $\msbm{R}^{9,1}$.
In this limit, the $U(N)$ symmetry is generically broken down to 
$U(1)^N$ with the Weyl group acting on the eigenvalues. 
We will refer to this limit as the broken phase.
We should note that all we said above are exactly the properties
of the ADHM description \ADHM\ of Yang-Mills instantons.

\subsec{Equivariant Cohomology}

In the space of matrices $\overline W$ the most natural object
is the $U(N)$ equivariant cohomology.
We introduce a generator $Q_+$ of the $U(N)$
equivariant cohomology on $\overline W$ satisfying
\eqn\eaa{
Q_+^2 = -i\d_{\phi_{++}},
}
where $\d_{\phi_{++}}$ denote $U(N)$ transformation generated
by $\phi_{++}$, 
which is a $N\times N$ matrix in the adjoint representation
of $U(N)$. We have the basic action of the algebra
\eqn\eab{
Q_+ X^\m = i\p_+^\m,\qquad Q_+ \p_+^\m = -[\phi_{++}, X^\m],
\qquad Q_+ \phi_{++}=0,
}
where $\p^\m_+$ is a $N\times N$ matrix with anti-commuting
matrix elements.
We define an additive quantum number $U$ and assign
$U=1$ to $Q_+$. We restrict to the $U(N)$-invariant subspace 
by setting
$Q^2_{++}=0$, which reduces to ordinary cohomology provided 
that $U(N)$ acts freely.
More physically we can interpret 
the transformation law $Q_+X^\m =i\p^\m_+$
as the BRST-like symmetry for the invariance
under the arbitrary shift $X^\m\rightarrow X^\m +\d X^\m$.
Thus $\p^\m_+$ is nothing but a ghost.
The second transformation law in \eab\ involves
the redundancy of our description.
The general idea of TFT is to study a certain moduli
problem using the action functional constructed by gauge fixing
the symmetry denoted by $Q_+$. The moduli space
is defined by the solution space, modulo gauge symmetry,
of certain field equations (matrices in our case).
Then, $\p^\m_+$ is required to satisfy certain linearized
equation as well as to be orthogonal to the direction
of the $U(N)$ rotation.

We can extend our equivariant cohomology to its balanced
version \DM. In the balanced equivariant cohomology
one introduces another fermionic charge $Q_-$
carrying $U=-1$ and the corresponding copy of \eab,
\eqn\eac{
Q_- X^\m = i\p_-^\m,\qquad Q_- \p_-^\m = -[\phi_{--}, X^\m],\qquad
Q_- \phi_{--}=0,
}
satisfying
\eqn\ead{
Q_-^2 = -i\d_{\phi_{--}}.
}
To make the algebra of our system complete we have to
decide about the mutual commutation relation between
the two generators $Q_\pm$.
The simplest possibility might be
$\{Q_+,Q_-\}=0$. This choice however is inconsistent. 
Thus we are led to introduce another generator
of the $U(N)$ symmetry which has $U=0$. 
We have to introduce a new matrix $\phi_{+-}$
and postulate
\eqn\eae{
\{Q_+,Q_-\} = -i2\d_{\phi_{+-}}.
}
The commutation relations \eaa\ and \ead\ together with \eae\
determine the superalgebra in a {\it unique} way.
Note that the three
separate $U(N)$ symmetry generators 
$(\phi_{++},\phi_{+-},\phi_{--})$ carry
$U=(2,0,-2)$. We will usually denote $\phi_{++}=\phi$,
$\phi_{+-}=C$
and $\phi_{--}=\bar\phi$.
To complete the action of the generators $Q_\pm$ we further 
have to introduce auxiliary matrices $H^\m$. They are introduced
 in the algebra as
\eqn\eaf{
\eqalign{
&Q_+ \p^\m_- = +H^\m - [C, X^\m],\cr
&Q_- \p^\m_+ = -H^\m - [C,X^\m],\cr
}
}
which agrees with \eae, i.e., $\{Q_+,Q_-\}X^\m = -2i[C,X^\m]$.
To make the algebra closed, we need to impose the
following consistent conditions
\eqn\eag{
\eqalign{
Q_+ \bar\phi + 2 Q_-C=0,\cr
Q_- \phi + 2 Q_+ C=0,\cr
}\qquad
\eqalign{
Q_+^2 \bar\phi = -i[\phi,\bar\phi],\cr
Q_-^2 \phi = -i[\bar\phi,\phi],\cr
}
\qquad
\{Q_+, Q_-\} C =0.
}
These may be seen as the Jacobi identities of the algebra. 
The solution is
\eqn\eah{
\eqalign{
&Q_+C=i\xi_+,\cr
&Q_-C=i\xi_-,\cr
}
\qquad
\eqalign{
&Q_+ \bar\phi =-{2}i\xi_-,\cr
&Q_- \phi =-{2}i\xi_+,\cr
}\qquad
\eqalign{
&Q_+\xi_- = +\Fr{1}{2}[\phi,\bar\phi],\cr
&Q_-\xi_+ = -\Fr{1}{2}[\phi,\bar\phi],\cr
}\qquad
\eqalign{
&Q_+\xi_+ = -[\phi, C],\cr
&Q_-\xi_- = -[\bar\phi,C],\cr
}
}
Finally consistency with the algebra leads to a transformation of 
the auxiliary fields $H^\m$ given by
\eqn\eaj{\eqalign{
&Q_+ H^\m = -i[\phi,\p^\m_-] +i[C,\p^\m_+] +i[\xi_+, X^\m],\cr
&Q_- H^\m = +i[\bar\phi,\p^\m_+] -i[C,\p^\m_-] -i[\xi_-, X^\n],\cr
}
}
One can check that the algebra is closed.

Before proceeding we summarize the contents of our matrices.
We have ten commuting matrices $X^\m$ and their
fermionic partners $\p^\m_+$ and $\p^\m_-$,
with $U=1$ and $U=-1$ respectively They carry an $SO(9,1)$
vector index $\m=0,\ldots,9$. We have $10$ bosonic auxiliary
matrices $H^\m$ carrying $U=0$ and an
$SO(9,1)$ vector index. 
We also have
three bosonic matrices $\phi$, $C$ and $\bar\phi$ carrying $U=2$, 
$U=0$ and 
$U=-2$ respectively. Those matrices have superpartners
$\xi_+$ and $\xi_-$ with $U=1$ and $U=-1$, respectively.
Note that our algebra has an internal $sl_2$ structure \VW\DM.
The matrices $X^\m$ form ten copies of an $sl_2$ singlet,
$(\p_+^\m, \p_-^\m)$ form ten copies of an $sl_2$ doublet,
$(\phi,C, \bar\phi)$ form an $sl_2$ triplet
and $(\xi_+,\xi_-)$ form an $sl_2$ doublet.
All this can be nicely summarized by the following
diagram \DM
\eqn\ymvi{
\matrix{
U=+2\cr
{}\cr
U=+1\cr
{}\cr
U=0\cr
{}\cr
U=-1\cr
{}\cr
U=-2\cr
}\qquad
{\rm fields}\quad
\matrix{    &  &   \p^\m_{+}  &  &  \cr
    & \nearrow &  &  \searrow & \cr
X^\m &  & & & H^\m\cr
 & \searrow & & \nearrow & \cr
 & & \p^\m_{-}& & \cr},
\qquad {\rm consistency}\quad
\matrix{
\phi_{++}&          &      \cr
         &\searrow  &      \cr
         &          &\xi_+\cr
         &\nearrow  &      \cr
\phi_{+-}&          &      \cr
         &\searrow  &      \cr
         &          &\xi_-\cr
         &\nearrow  &      \cr
\phi_{--}&          &      \cr
}
}
The $sl_2$ symmetry of our algebra is referred to as the balanced 
structure. The symmetry under filliping the signs of the $U$-number
implies that the net $U$-number of fermionic zero-modes
is always zero.  We will refer the first multiplet to
a vector multiplet.

\subsec{Action Functional}

Now we have enough machinery to
define the
action functional, 
which should have $SO(9,1)\times U(N)$ symmetry and is invariant
under the
$\CN=2$ symmetry generated by $Q_\pm$. 
As a BTFT we also require the action functional
to be invariant under the $sl_2$ symmetry. In particular, 
the action functional
should have $U=0$. The desired action functional turns out to be
almost {\it uniquely} determined.\foot{This
is a general property of BTFT \DM.} To begin with
we define
\eqn\mac{
S_1 = Q_+Q_-\CF_1,
}
derived of a supersymmetry transformation of the action potential
\eqn\mad{
\CF_1
= -\tr\left(2\p^\m_+ \p_{\m-}
+ \xi_-\xi_+\right).
}
Here $\CF_1$ is uniquely determined by the global $SO(9,1)$ 
and $sl_2$
symmetries.
We find
\eqn\vvg{
\eqalign{
S_1 = 
&\tr\biggl(
2[\phi, X^\m][\bar\phi, X^\n]
+2i\p^{\m}_-[\phi,\p_{\m-}]
+2i\p^{\m}_+[\bar\phi,\p_{\m+}]
+4i[C,\p^\m_+]\p_{\m-}
\cr
&\phantom{\biggl(}
+4i[X^\m,\p_{\m+}]\xi_-
+4i[X^\m,\p_{\m-}]\xi_+
-2[C,X^\m][C,X_\m]
+2H^\m H_\m
\cr
&\phantom{\biggl(}
-[\phi,C][\bar\phi,C]
-i\xi_-[\phi,\xi_-]
-i\xi_+[\bar\phi,\xi_+]
+2i\xi_+[C,\xi_-]
-\Fr{1}{4}[\phi,\bar\phi]^2
\biggr).
}
}

For our purpose the above action functional is not good enough.
We need to generate a potential term $V=[X^\m,X^\n]^2$ for 
the $X^\m$ such that these matrices
commute in the flat direction. To get this term we need a 
cubic action potential term $\CF_0$. However there are no $sl_2$
and $SO(9,1)$ invariant combinations
of the existing matrices $X^\m$ such that $Q_+Q_-\CF_0$ generates 
this potential. Consequently we have to introduce one more
matrix multiplet. We introduce a new adjoint matrix $B^{\m\n}$
carrying $U=0$ which is
anti-symmetric in the $SO(9,1)$ indices.
We have a corresponding algebra
\eqn\newa{
\eqalign{
&Q_+ B^{\m\n} = i\chi^{\m\n}_+,\cr
&Q_- B^{\m\n} = i\chi^{\m\n}_-,\cr
}\qquad
\eqalign{
&Q_+ \chi^{\m\n}_+ = -[\phi, B^{\m\n}],\cr
&Q_+ \chi^{\m\n}_- = + H^{\m\n} - [C,B^{\m\n}],\cr
&Q_- \chi^{\m\n}_+ = - H^{\m\n} - [C,B^{\m\n}],\cr
&Q_- \chi^{\m\n}_- = -[\bar\phi, B^{\m\n}],\cr
}
}
We will refer to  the above multiplet as the anti-symmetric
tensor multiplet.
We define
\eqn\eert{
S_0 + S_2 = Q_+Q_-\biggl(\CF_0 +\CF_2\biggr),
}
with
\eqn\deft{
\CF_0 = -\tr\left( iB^{\m\n}\left([X_\m,X_\n]
+\Fr{1}{3} [B_{\m\r}, B_{\n}{}^\r]
\right)\right),\qquad
\CF_2 =-\tr \left(\chi^{\m\n}_+ \chi_{\m\n -}\right),
}
which are again the only two $sl_2$ and $SO(9,1)$ invariants,
which do not introduce bare mass.\foot{There are other 
combinations,
which are redundant. We will consider the massive deformations
in a later section.}

Working through the algebra, we obtain the complete action 
\eqn\maf{
\eqalign{
S_0 + S_2 = 
&\tr\biggl(
[\phi,B^{\m\n}][\bar\phi,B^{\m\n}]
+i\chi^{\m\n}_-[\phi,\chi_{\m\n -}]
+i\chi^{\m\n}_+[\bar\phi,\chi_{\m\n +}]
+2i[C,\chi^{\m\n}_+]\chi_{\m\n-}
\cr
&\phantom{\biggl(}
+2i [B_{\m\n},\chi^{\m\n}_{+}]\xi_-
+2i [B_{\m\n},\chi^{\m\n}_-]\xi_+
-[C,B^{\m\n}][C,B_{\m\n}]
+H^{\m\n}H_{\m\n}
\cr
&\phantom{\biggl(}
-H^{\m\n}\bigl([X_\m,X_\n]+[B_{\m\r},B_\n{}^\r]\bigr)
+2H^\m[B_{\m\n},X^\n]
-2i B_{\m\n}[\p^\m_+,\p^\n_-]
\cr
&\phantom{\biggl(}
-2i B^{\m\n}[\chi_{\m\r_+},\chi_\n{}\r]
+2i\chi^{\m\n}_-[X_\m,\p_{\n +}]
-2i\chi^{\m\n}_+[X_\m,\p_{\n -}]
\biggr).
}
}
Now we define the total action $S$ by
\eqn\tat{
S= S_0 + S_1 + S_2.
}
We can integrate out the auxiliary matrices $H_\m$ and $H_{\m\n}$
by setting
\eqn\repl{
H_\m = -\Fr{1}{2}[B_{\m\n},X^\n],
\qquad
H_{\m\n} = \Fr{1}{2}[X_\m,X_\n] +\Fr{1}{2}[B_{\m\r},B_\n{}^\r]
}
to get
\eqn\maff{
\eqalign{
S&= 
\tr\biggl(
2[\phi, X^\m][\bar\phi, X^\n]
+2i\p^{\m}_-[\phi,\p_{\m-}]
+2i\p^{\m}_+[\bar\phi,\p_{\m+}]
+4i[C,\p^\m_+]\p_{\m-}
\cr
&\phantom{\biggl(}
+4i[X^\m,\p_{\m+}]\xi_-
+4i[X^\m,\p_{\m-}]\xi_+
-2[C,X^\m][C,X_\m]
-[\phi,C][\bar\phi,C]
\cr
&\phantom{\biggl(}
+2i\xi_+[C,\xi_-]
-i\xi_-[\phi,\xi_-]
-i\xi_+[\bar\phi,\xi_+]
-\Fr{1}{4}[\phi,\bar\phi]^2
+[\phi,B^{\m\n}][\bar\phi,B_{\m\n}]
\cr
&\phantom{\biggl(}
+2i[C,\chi^{\m\n}_+]\chi_{\m\n-}
+i\chi^{\m\n}_-[\phi,\chi_{\m\n -}]
+i\chi^{\m\n}_+[\bar\phi,\chi_{\m\n +}]
+2i [B_{\m\n},\chi^{\m\n}_{+}]\xi_-
\cr
&\phantom{\biggl(}
+2i [B_{\m\n},\chi^{\m\n}_-]\xi_+
-2i B_{\m\n}[\p^\m_+,\p^\n_-]
-2i B^{\m\n}[\chi_{\m\r_+},\chi_\n{}^\r_-]
\cr
&\phantom{\biggl(}
+2i\chi^{\m\n}_-[X_\m,\p_{\n +}]
-2i\chi^{\m\n}_+[X_\m,\p_{\n -}]
-[C,B^{\m\n}][C,B_{\m\n}]
\cr
&\phantom{\biggl(}
-\Fr{1}{4}\left([X_\m,X_\n] 
+\Fr{1}{4}[B_{\m\r},B_\n{}^\r]\right)^2
-\Fr{1}{2}[B^{\m\r},X_\n]^2
\biggr).
}
}

This action is invariant under the $Q_\pm$ symmetries 
after replacing
$H^\m$ in \eaf\ with the expression in \repl.
As a TFT, we study the fixed points of $Q_\pm$ symmetry.
First of all, from \eaf. \newa\ and \repl, the fixed
point equations $Q_\pm \p^\m_\mp=0$ and $Q_\pm \chi^{\m\n}_{\mp}=0$
imply $H_\m=H_{\m\n}=0$, which is equivalent to
\eqn\localization{
[X_\m,X_\n] +\Fr{1}{4}[B_{\m\r},B_\n{}^\r]=0,
\qquad
[B^{\m\n},X_\n]=0,
}
and
\eqn\strafi{
\qquad
[C,X_\m]=[C, B_{\m\n}]=0.
}
We also have other fixed point equations
\eqn\ffr{
[\phi_{AB},\phi_{A^\pr B^\pr}]=0,\qquad 
[\phi_{AB},X^\m]=[\phi_{AB},B^{\m\n}]=0.
}
These are the equations for the localization which determine
the moduli space we want to study. The $\chi^{\m\n}_\mp$,
$\p^\m_{\mp}$ and
$\xi_\mp$ equations of motion, modulo the $U(N)$ symmetry
generated by $\phi, C,\bar\phi$ are
\eqn\linco{
[X_\m,\p_{\pm\n}] +[B_{\m\r},\chi_{\n\pm}{}^\r]=0,\quad
[\chi^{\m\n}_\pm,X_\n] +[B^{\m\n},\p_{\pm\n}]=0,
\quad
\eta_{\m\n}[X^\m,\p^\n_\pm]
+ [B^{\m\n}, \chi_{\m\n\pm}]=0.
}
The first two equations can be interpreted as the linearization of 
\localization\ and the last equation can be interpreted
as a kind of Coulomb gauge fixing condition.

We define the partition function $Z_0$ by
\eqn\mag{
Z_0 = \Fr{1}{\hbox{\it Vol}(G)}
\int_{\overline\CW\otimes (Lie(\CG)\otimes Lie(\CG)^*)}
\!\!\!\!\!\!\!\!\!\!\!\!\!\!\!\!\!\!\!\!\!\!\!\!\!
\CD X\CD \p_+ \CD \p_- 
\CD B \CD\chi_+\CD\chi_-
\CD \phi\CD \bar\phi \CD C \CD\xi_+\CD\xi_-
\times e^{-S_0}.
}
Note that we are dealing with a topological theory
so that the stationary phase evaluation is
exact. In other words the path integral is localized
to the space of
supersymmetric minima
of the action given by \localization.
At this point we like to emphasis the distinction
between $X^\m$ and $\phi_{AB}$. Note that we introduced
$\phi_{AB}$ as the generators of $U(N)$ symmetry of the
matrices $X^\m$. In other words the matrices
$\phi_{AB}$ are responsible for
 {\it pure gauge degrees of freedom}.
So the equations $[\phi_{AB},\phi_{A^\pr B^\pr}]=0$ 
define the {\it flat} directions.
We can diagonalize $\phi_{AB}$ simultaneously. Then the flat
direction can be identified with $Sym^{N}(\msbm{R}^3)$.
Now we see that, from \strafi, 
the supersymmetric
minimum configuration is a {\it multiply stratified} 
space
parameterized by a point in $Sym^{N}(\msbm{R}^3)$.
Note that the supersymmetric minimum depends only on
a particular stratum of $Sym^{N}(\msbm{R}^3)$ determined
by the symmetry breaking pattern of $U(N)$.
At generic points in $Sym^{N}(\msbm{R}^3)$ the $U(N)$ symmetry
is broken down to $U(1)^N$. In a diagonal some non-Abelian
symmetry is restored. 

The simplest solutions to \localization\
are given by the case where the ten matrices $X^\m$ are
mutually commuting and $B^{\m\n}=0$.
So $\{X^\m\}$ can be simultaneously diagonalized.
Such a diagonalization depends on a point in $Sym^N(\msbm{R}^3)$.
We can interpret the eigenvalues $x^\m_{\ell}$, $\ell=1,\ldots,N$,
as the positions of $N$ unordered points in a {\it space-time}
$\msbm{R}^{9,1}$. In other words, we are describing a system
of $N$ point-like instantons in ten-dimensional 
Minkowski space-time
as the supersymmetric minimum.  Now our abstract global symmetry
group $SO(9,1)$ can be interpreted as the Lorentz symmetry
of $\msbm{R}^{9,1}$.

How about more general solutions of \localization\ ?
For example, we can imagine solutions with non-vanishing
$B^{\m\n}$, either commuting or non-commuting one.
For non-commuting $B^{\m\n}$, leading to non-commuting
$X_\m$, we may use some analogy with the monads (ADHM) 
construction
of Yang-Mills instantons. Those degrees of freedom 
may be attributed
to the size and relative degrees among instantons. 
For commuting and non-trivial solutions of $B_{\m\n}$
we certainly have problems in the space-time interpretations.
Furthermore, we can allow more general solutions which
break our $SO(9,1)$ symmetry. Then some components
of $B_{\m\n}$ can be interpreted as positions of
instantons living in the lower dimensional space.
Such new matrices transform as vectors under
the smaller Lorentz group defining another ``non-commuting''
space-time coordinates of instantons. 
We will refer to all those solutions
as {\it almost universal instantons}.
The systems we are describing
can be interpreted as {\it monads} of such instantons which
we will refer to as {\it almost universal monads}.\foot{
Note that above spacetime interpretation are motivated from
the ADHM description of Yang-Mills instantons as well as
Witten's description of D-instantons (D-branes in general).
Witten also mentioned the intriguing similarity between
the two cases. This observation is, actually, the starting
point of our investigation.}

There are many other issues concerning the model constructed
in the section. Since we will have to repeat those
in  our description of monadic string, we will not discuss them
here. But we like to clarify  
the role of the anti-symmetric tensor $B^{\m\n}$. 
It was not entirely clear, in the treatment of this section, 
how we can 
interpret the eigenvalues of $B^{\m\n}$. However, we had
to introduce $B^{\m\n}$ to define a meaningful theory.
Note that $B^{\m\n}$ was introduced because of the non-commutative
nature of ``spacetime'' coordinates of instantons
and the requirement of{\it covariance}. Thus we can 
naturally expect that the existence of ``spacetime coordinates''
as antisymmetric tensors may be just the direct
requirement for the covariant description
of  the existence
of off-diagonal parts of ``spacetime coordinates''. 
In the next
section, we will discuss these issues for the similar
description of instantons in four-dimensions.
In later sections, we will return to those points again.

\newsec{Extended Monads and $\CN=4$ SYM Theory in Four-Dimensions}

In this section we will consider a system of four matrices
$X^i$, $i=0,1,2,3$ rather than ten matrices.
To relate with Yang-Mills instantons we assume the  $X^i$ to
transform as the components of a vector for $SO(4)$. 
We will repeat the construction
of the previous section in the new setting. We will find
relations with the monads (the ADHM) description
of Yang-Mills instantons. We will discuss the interpretation
of ``space-time coordinates'' which transform as
tensors or scalars. We  also discuss close relations
with the Vafa-Witten model of twisted $\CN=4$ SYM theory 
(or BTYMT) on a four-manifold \VW\DM. 
Using the structure of BTYMT, 
we will recall Witten's arguments on the unbroken phase of quantum 
gravity.

\subsec{A Description of Instantons 
in four-dimensions}

This sub-section can be viewed as a continuation
of the paper \Monads, where the monads (the ADHM equation)
construction
of Yang-Mills instanton was extended in a way motivated by
the Vafa-Witten equation of $\CN=4$ SYM theory and 
its relation with the Seiberg-Witten equation.
The equations of extended monads are simply the reduction
of the Vafa-Witten equations to zero-dimensions.
We can repeat the
same constructions as in the two previous subsections.

It is possible to break half the supersymmetry
maintaining only the symmetry generated by $Q_+$. 
An important perturbation satisfying this constraint 
is given by adding bare mass terms with non-zero 
$U$-number to the action. Since the theory in the bulk
is $U$-number anomaly free, such a perturbation does
not change the theory unless we take a very special limit.
We may view such perturbations as looking to the system through
a magnifying glass.
Essentially the same perturbation is discussed in \VW\ and
\DPS.
The resulting theory will be
localized to the fixed point locus of this $Q_+$ symmetry
given by
\eqn\vawt{\eqalign{
\Fr{1}{2}[X^i,X^j] 
+\Fr{1}{2}[B^{i\ell}, B^{j}{}_\ell]
-[C,B^{ij}]=0,\cr
[X_i,B^{ij}] + [X^j,C]=0.
}
} 
supplemented by the equations
\eqn\fdsa{
[\phi,\bar\phi]=0,\qquad
[\phi, T_i]=0.
}
We can decompose the anti-symmetric tensor $B_{ij}$ 
under $SO(4)$ into its self-dual and the anti-self-dual
parts
\eqn\decomps{
B_{ij}= B^+_{ij} + B^-_{ij},
}
and the two components are orthogonal to each others.
Now we can consider the self-dual part of the
the equations \vawt
\eqn\vafawittenz{
\eqalign{
\Fr{1}{2}[X^i,X^j]^+ 
+\Fr{1}{2}[B^{+ i\ell}, B^{+ j}{}_{\ell}]
-[C,B^{+ ij}]=0,
\cr
[X_i,B^{+ ij}] + [X^j,C]=0.
}
} 
The above self-dual truncation is nothing but
the Vafa-Witten equations reduced
to zero-dimensions \VW. In the paper \Monads,
we referred to them as the equations for extended monads.
By using complex $SO(4)$ indices we can rewrite
the equations \vafawittenz\ as equations for $4$ complex
$N\times N$ matrices $T_a$, $a=1,\ldots,4$;
\eqn\exmon{\eqalign{
&[T_1,T_2] +[T_3,T_4]=0,\cr
&[T_1,T_1^*]+[T_2,T_2^*] +[T_3,T_4]=0,
}
}
where $T_1$ and $T_2$ are build out of the $X^i$ and
$T_3$ and $T_4$ come from $(C, B^+_{ij})$.
We only wrote down the first of the equations
\vafawittenz. We certainly have solutions of \exmon\
with all of the $T_a$ simultaneously diagonalized,
$U T_a U^{-1}= t_a$, where the $t_a$ are diagonal matrices. 
We might interpret the
eigenvalues $t_i^\ell$ of $t_i$ as the positions of $N$
points in $\msbm{R}^8$. There is however an 
obvious problem to such an interpretation
 since $T_3$ and $T_4$ do not
transform as the components of a vector for $SO(4)$. 
Note that we can still
interpret the eigenvalues of $T_1$ and $T_2$
as the positions of points in $\msbm{R}^4$.
In fact, if we set $T_3$ and $T_4$ to zero, the equation 
\exmon\ is nothing but the ADHM equations of $N$ point-like
(Yang-Mills) instantons in $\msbm{R}^4$.

A solution to the problem above was presented in the paper \Monads.
In \Monads\ and \DPS, the breaking of $Q_\pm$ to $Q_+$ was
realized by extending the Dolbeault version of the
balanced equivariant cohomology to incorporate
the obvious global symmetry
\eqn\soaction{
(T_3,T_4) \rightarrow (e^{-im\th}T_3, e^{im\th}T_4).
}
As a result, the fixed point equations \fdsa\ should be
modified to
\eqn\modr{
[\phi,\bar\phi]=0,\qquad
[\phi, T_1]=[\phi,T_2]=0,
\qquad
\eqalign{
[\phi,T_3]=+m T_3,\cr
[\phi,T_4]=-m T_4,\cr
}
}
where $m$ is the bare mass. Now it is obvious that
there are no non-trivial diagonal solutions for $T_3$ and $T_4$
for $m\neq 0$.
Their solutions are always off-diagonal 
so that we will never be able
to interpret them as positions or coordinates in space-time!
The situation was described in detail in the paper \Monads.
We will recall two typical solutions. The solutions of
\exmon\ and \modr\ are determined by the symmetry breaking
pattern of $U(N)$ (via the eigenvalues of $\phi$).
If the $U(N)$ symmetry is unbroken $T_3=T_4=0$ and
$T_1$ and $T_2$ should be simultaneously diagonalized.
Then we get the ADHM description of $N$ point-like instantons
in $\msbm{R}^4$. If the $U(N)$ symmetry is broken down to
$U(N-k)\times U(k)$ we find
\eqn\waste{
T_1=\left(\matrix{t_1&0\cr 0&t_1^\pr}\right),
\qquad
T_2=\left(\matrix{t_2&0\cr 0&t_2^\pr}\right)
,\qquad
T_3 =\left(\matrix{0&\s\cr 0&0}\right)
,\qquad
T_4 =\left(\matrix{0&0\cr \pi&0}\right),
}
where 
$\s$ is $k\times (N-k)$ and $\pi$ 
is $(N-k)\times k$ matrices.
We have
\eqn\xxa{
\left\{\eqalign{
&[t_1, t_2] + \s\pi=0,\cr
&[t_1,t_1^*] +[t_2,t_2^*] + \s\s^* 
-\pi^*\pi=0,
}\right.\quad
\left\{\eqalign{
&[t_1^\pr, t_2^\pr] -\pi\s=0,\cr
&[t_1^\pr, t_1^{\pr *}] +[t_2^\pr,t_2^{\pr *}] 
 +\pi\pi^*-\s^*\s=0.\cr
}\right.
}
Note that the first and the second set of equations describe
$SU(N-k)$ and $SU(k)$ Yang-Mills instantons with instanton
numbers $k$ and $(N-k)$ respectively.
Now the role of $T_3$ and $T_4$ is clear. They carry information
about the gauge group and the size of Yang-Mills instantons
in $\msbm{R}^4$.

In the above discussions we restrict our attention to
the self-dual part $B^+_{ij}$ of $B_{ij}$. This restriction
can easily be justified. Recall that $B_{ij}$ is introduced
to get the crucial potential term $\tr[X_i,X_j][X^i,X^j]$.
We can decompose $[X_i,X_j]$ into  self-dual and
anti-self-dual parts and show that
\eqn\ggg{
\tr[X_i,X_j][X^i,X^j]= 2 \tr[X_i,X_j]^+[X^i,X^j]^+.
}
Thus the anti-self-dual part $B_{ij}^-$ of $B_{ij}$
is redundant. This implies that we are describing
essentially the same system with the self-dual 
anti-symmetric tensor multiplet only.

\subsec{The Global $\CN=4$ Super-Yang-Mills Theory}

Now we consider $\CN=4$ super-Yang-Mills theory \VW\DM.
Let $M$ be an arbitrary four-manifold where our $SO(4)$ symmetry
is acting. Let $E$ be a $U(N)$ bundle over $M$ and
let $X^i$ be the components of a connection. The BTYM theory is
defined exactly as in Sect.~$2.1$ with the same
commutation relations \eaa. The only change
is that the $U(N)$ gauge transformation acts on $X^i$
by
\eqn\werar{
X_i \rightarrow g X_i g^{-1} + g \rd_i g^{-1},
}
where $g :M\rightarrow U(N)$. In the space of all connections
$\CX$ we have a natural metric
\eqn\metrsp{
\left|\d X\right|^2=\int_M d\m\tr\left( \d X^i \d X_i\right),
}
where $d\m$ denotes the measure on $M$.
Every other field transforms in the adjoint representation.
The algebra is given by
\eqn\basics{
Q_\pm X^i = i\p^i_\pm,\qquad
Q_\pm \p^i_\pm = D_i \phi_{\pm\pm},
\qquad
Q_\pm \p^i_{\mp}=\pm H^i + D^i C,
}
where $D_i$ is the gauge covariant derivative. 
The remaining algebra is left unchanged.

The global 
$\CN=4$ (space-time) supersymmetry requires that the anti-symmetric
tensor multiplet is self-dual.
Apart from the underlying space-time supersymmetry,
as in the previous subsection,
the restriction to a self-dual anti-symmetric tensor multiplet
is a very natural requirement. Now the
potential term becomes the usual kinetic term
$\tr F\wedge * F$ of
Yang-Mills theory. The well-known fact that
\eqn\yma{
\int_M\tr F\wedge * F = 2\int_M \tr\left( F^+\wedge * F^+\right)
+8\pi^2 k,
}
where $k$ denotes the instanton number
\eqn\instasn{
k= \Fr{1}{8\pi^2}\int_M \tr\left(F\wedge F\right),
}
implies that it is sufficient to introduce the self-dual
part of the anti-symmetric tensor multiplet. We can freely add
the topological term \instasn\ to our action without spoiling
anything.
The action functional is defined by \DM
\eqn\dfg{
S = \Fr{1}{e^2}Q_+Q_-\int_M d\m
\biggl(\CF_0^+ + \CF_1 +\CF_2^+\biggr),
}
where the $\CF$'s are given as in \deft\ and the superscript $+$
denote that we only use the self-dual part of the
anti-symmetric tensors. Clearly $[X_i,X_j]$ in $\CF_0$
should be replaced with the field strength 
$F_{ij}$.
Here $e^2$ denotes the Yang-Mills coupling constant
which are dimensionless.

\newsec{Monad String Theory}

In the previous section we extended the monad (ADHM) description
of $N$ point-like instantons in $\msbm{R}^4$ to $\msbm{R}^{9,1}$
and construct, presumably the most natural, supersymmetric
theory out of it. 
In this section we will apply the same
ideas to describe second quantized superstring theory in 
$\msbm{R}^{9,1}$. Throughout this section we will restrict
our attentions to classical aspects of the model.

\subsec{The Algebra and Action Functional}

To begin with we assume
our ten matrices $X^\m$ to be matrix functions $X^\m(\s^+,\s^-)$
of two world-sheet coordinates. Let $\overline W(\s^+,\s^-)$
be the space of $N\times N$ Hermitian matrix functions.
We endow the space $\overline W(\s^+,\s^-)$ with the natural metric
\eqn\eaa{
\left|\d X\right|^2 = 
\int d\s^+ d\s^-\tr\left(\eta_{\m\n} \d X^\m \d X^\n\right).
}
This metric is invariant under local $U(N)$ symmetry
$X^\m\rightarrow g X^\m g^{-1}$ for $g\in \CG$
such that $g:\S\rightarrow U(N)$, where 
$\S$ denotes the ``world-sheet'' which is the space of parameters 
$\s^\pm$. As mentioned in the introduction we want to
construct a theory by gauge fixing the ``spacetime''
and ``world-sheet'' Poincar\'{e} symmetry as well as
the local $U(N)$ gauge symmetry. By the ``spacetime''
Poincar\'e symmetry, we mean the invariance under
the global $SO(9,1)$ symmetry acting on the ``spacetime''
vector index $\m$ and the invariance under arbitrary
shift $X^\m \rightarrow X^\m +\d X^\m$. Clearly they
are symmetries of our metric \eaa. From the viewpoint
of two-dimensional $U(N)$ gauge theory, the ``spacetime''
Lorentz covariance is just a global symmetry among
the fields $X^\m$. The ``spacetime'' translation invariance
implies that the two-dimensional gauge theory is a
TFT.

To get define a system with these properties we can simply extend our  balanced
$\CG$-equivariant cohomology to include translations along
the internal directions. We define the commutation relations
\eqn\mba{\eqalign{
&Q_+^2 = -i\left(\Fr{\rd}{\rd\s^+} +\d_{\phi_{++}}\right),\cr
&Q_-^2 = -i\left(\Fr{\rd}{\rd\s^-} +\d_{\phi_{--}}\right).\cr
}\qquad
\{Q_+,Q_-\}= -i \d_{\phi_{+-}}.
}
This immediately leads to the following
basic algebra
\eqn\mca{
\eqalign{
&Q_+ X^\m =i\p^\m_+,\cr
&Q_- X^\m =i\p_-^\m,\cr
}\qquad
\eqalign{
&Q_+ \p^\m_+ = -D_+X^\m,\cr
&Q_- \p^\m_- = -D_-X^\m.\cr
}
}
The above extension
is indeed a very natural step. For
$Q_\pm X^\m = i\p^\m_\pm$, we can interpret $\p^\m_\pm$ as
the ghosts for the topological symmetry of the arbitrary
shift $X^\m \rightarrow X^\m +\d X^\m$. This description
clearly has a redundancy which is the $U(N)$ symmetry and
the shift of parameters $\s^\pm$ as indicated in \mba.
We may interpret $Q_\pm$ as the balanced equivariant cohomology
generators of $U(N)\times P_{\s^\pm}$. We will see shortly
that the explicit realization of \mba\ requires that the
canonical line bundle of $\S$ is trivial. Naturally,
we will consider $\S$ to be a two-dimensional cylinder. 
This fits nicely with the description of closed string.
We can identify $\s^\pm$ with ``world-sheet'' light-cone 
coordinates, i.e., $\s^\pm=\Fr{1}{2}(\s \pm \t)$.
For consistency we see from \mba\ that $\phi_{\pm\pm}$ should 
transform as the components of an
$U(N)$ connection. So we can identify $\phi_{++}$
and $\phi_{--}$ with  the left and right components
of the $U(N)$ connection. The global $sl_2$ symmetry 
of the balanced
equivariant cohomology can be identified with
the ``world-sheet'' Lorentz symmetry. 
To put it differently, we are just extending the 
world-sheet supersymmetry to include the $U(N)$ symmetry -
the $\CG$-equivariant extension of world-sheet supersymmetry.

>From the construction in the previous section, it
is straightforward to get the modified algebra.
The algebra \mca\ is supplemented by
\eqn\mcc{\eqalign{
&Q_+ \p^\m_- = +H^\m -[C,X^\m],\cr
&Q_- \p^\m_+ = -H^\m -[C,X^\m],\cr
}
}
and
the algebra of consistency \eah\ should be
modified to
\eqn\mcb{
\eqalign{
&Q_+ C=i\xi_+,\cr
&Q_- C=i\xi_-.\cr
}\qquad
\eqalign{
&Q_+ \phi_{++}=0,\cr
&Q_- \phi_{++}= -2i\xi_+,\cr
&Q_+ \phi_{--}= -2i\xi_-,\cr
&Q_- \phi_{--}=0.\cr
}\qquad
\eqalign{
&Q_+\xi_+ = - D_+ C,\cr
&Q_+\xi_- = 
+\Fr{1}{2}F_{+-},\cr
&Q_-\xi_+ = 
-\Fr{1}{2}F_{+-},\cr
&Q_-\xi_- = -D_- C,\cr
}
}
where $F_{+-}$ is the Yang-Mills curvature of the $U(N)$ connection
$\phi_{\pm\pm}$.\foot{Under the local 
gauge transformation the connection $A$
transform as $\d_\e A = d_A\e$.}
Note also that the $U$ numbers of the covariant
derivatives $D_\pm$ are $\pm 2$. 
The triviality of the canonical
line bundle is required since we relate the ``world-sheet'' vector
$\phi_{\pm\pm}$ with a ``world-sheet'' scalar $C$ via \mcb.
The transformation laws for the auxiliary fields $H^\m$ are
\eqn\mcd{\eqalign{
&Q_+ H^\m = 
 -i D_+\p^\m_- 
 +i [C,\p^\m_+] 
 +i [\xi_+, X^\m],
 \cr
&Q_- H^\m = 
 +i D_-\p^\m_+
 -i[C,\p^\m_-] 
 -i[\xi_-, X^\m].\cr
}
}
A difference with the previous section is
that $(\phi_{++},\phi_{--})$ become an $sl_2$ doublet and
$C$ becomes an $sl_2$ singlet.

For reasons explained in the previous section
we introduce a multiplet $(B^{\m\n},\chi^{\m\n}_\pm,H^{\m\n})$
with the transformation laws
\eqn\nnw{
\eqalign{
&Q_+ B^{\m\n} = i\chi^{\m\n}_+,\cr
&Q_- B^{\m\n} = i\chi^{\m\n}_-,\cr
}\qquad
\eqalign{
&Q_+ \chi^{\m\n}_+ = -D_+ B^{\m\n},\cr
&Q_+ \chi^{\m\n}_- = + H^{\m\n} - [C,B^{\m\n}],\cr
&Q_- \chi^{\m\n}_+ = - H^{\m\n} - [C,B^{\m\n}],\cr
&Q_- \chi^{\m\n}_- = -D_-B^{\m\n},\cr
}
}
and
\eqn\nnww{\eqalign{
&Q_+ H^{\m\n}= -i D_+\chi^{\m\n}_- +i[C,\chi^{\m\n}_+] 
             +i[\xi_+, B^{\m\n}],\cr
&Q_- H^{\m\n}= +i D_-\chi^{\m\n}_+ -i[C,\chi^{\m\n}_-] 
             -i[\xi_-, B^{\m\n}].\cr
}
}
The scaling dimensions of 
$(B^{\m\n},\chi^{\m\n}_\pm, H^{\m\n})$ are
$(0,\fr{1}{2},1)$.

The action functional can be defined through 
a procedure similar to  Sect.~2. 
Only now we have to replace the $U(N)$ trace $\tr$ with
$\int d\s^+ d\s^-\tr$.  
The action functional can hence be written in the form
\eqn\dfg{
S = Q_+Q_-\left(\int d\s^+ d\s^-
\biggl(\CF_0 + \CF_1 + \CF_2\biggr)\right),
}
where the action potential terms are given by
\eqn\pppf{
\eqalign{
&\CF_0 =
-\tr\biggl(
iB^{\m\n}\left([X_\m,X_\n]+\Fr{R^2}{3}[B_{\m\r},B_{\n}{}^\s]
\right)\biggr),\cr
&\CF_1 = -\tr\biggl(2\p^\m_+ \p_{\m-} +R^2\xi_-\xi_+
\biggr),\cr
&\CF_2= -\tr\biggl(R^2
\chi^{\m\n}_+\chi_{\m\n-}\biggr).
}
}
We made a slightly more general choice for our action potential
than in the previous section by introducing a free parameter $R$.
Since our construction is the unique non-Abelian generalization
of the RNS string theory, we certainly expect to get
the free RNS string in a suitable limit, where the $U(N)$
symmetry breaks down to $U(1)^N$ and all fields can be
simultaneously diagonalized. Without introducing the free-parameter
$R$, we do not get the free RNS string.
Instead we get a superconformal theory consisting of both
the $X^\m$ and $B^{\m\n}$ multiplets.
Only after introducing $R$ and in the limit $R= 0$,
we get the free RNS string. 

Although this looks arbitrary, our choice is very natural since
the RNS string action entirely comes from the 
term $\tr(\p^\m_+\p_{\m-})$ in the action potential $\CF_1$.
We will see in a later section that the above
form of the action-potential originates from
eleven-dimensional covariance. It will become clear
that our construction is directing us to an underlying
theory with eleven dimensional covariance.

The explicit form of the action functional
is  
\eqn\trial{
\eqalign{
S&=\int d\s^+ d\s^- \tr\biggl(
2D_+X^\m D_-X_\m
+2i\p^{\m}_-D_+\p_{\m-}
+2i\p^{\m}_+D_-\p_{\m+}
+4i[C,\p^\m_+]\p_{\m-}
\cr
&\phantom{\biggl(}
+4i[X^\m,\p_{\m+}]\xi_-
+4i[X^\m,\p_{\m-}]\xi_+
-2[C,X^\m][C,X_\m]
+R^2 H^{\m\n}H_{\m\n}
+2H^\m H_\m
\cr
&\phantom{\biggl(}
-R^2 D_+ C D_- C
-iR^2 \xi_-D_+\xi_-
-iR^2 \xi_+D_-\xi_+
+2iR^2\xi_+[C,\xi_-]
-\Fr{R^2}{4}F_{+-}^2
\cr
&\phantom{\biggl(}
-H^{\m\n}\bigl([X_\m,X_\n]+R^2 [B_{\m\r},B_\n{}^\r]\bigr)
+2H_\m[B^{\m\n},X_\n]
+R^2 D_+ B^{\m\n} D_- B_{\m\n}
\cr
&\phantom{\biggl(}
+iR^2 \chi^{\m\n}_- D_+\chi_{\m\n -}
+iR^2 \chi^{\m\n}_+ D_-\chi_{\m\n +}
+2iR^2 [B_{\m\n},\chi^{\m\n}_{+}]\xi_-
+2iR^2 [B_{\m\n},\chi^{\m\n}_-]\xi_+
\cr
&\phantom{\biggl(}
+2iR^2[C,\chi^{\m\n}_+]\chi_{\m\n-}
-R^2[C,B^{\m\n}][C,B_{\m\n}]
-2iR^2B^{\m\n}[\chi_{\m\r+},\chi_{\n-}^{\phantom{-}\r}]
\cr
&\phantom{\biggl(}
-2i B_{\m\n}[\p^\m_+,\p^\n_-]
+2i\chi^{\m\n}_-[X_\m,\p_{\n +}]
-2i\chi^{\m\n}_+[X_\m,\p_{\n -}]
\biggr).
}
}
We can integrate out the auxiliary fields $H^{\m\n}$
and $H^\m$ by setting 
\eqn\reppl{
H_{\m\n} = \Fr{1}{2 R^2}
\left([X_\m,X_\n] +R^2[B_{\m\r},B_\n{}^\r]\right)
\qquad
H_\m = -[B_{\m\n},X^\n].
}
After this replacement we get
\eqn\ttrial{
\eqalign{
S^\pr&=\int d\s^+ d\s^- \tr\biggl(
2D_+X^\m D_-X_\m
+2i\p^{\m}_-D_+\p_{\m-}
+2i\p^{\m}_+D_-\p_{\m+}
+4i[C,\p^\m_+]\p_{\m-}
\cr
&\phantom{\biggl(}
+4i[X^\m,\p_{\m+}]\xi_-
+4i[X^\m,\p_{\m-}]\xi_+
-2[C,X^\m][C,X_\m]
\cr
&\phantom{\biggl(}
-R^2 D_+ C D_- C
-iR^2 \xi_-D_+\xi_-
-iR^2 \xi_+D_-\xi_+
+2iR^2\xi_+[C,\xi_-]
-\Fr{R^2}{4}F_{+-}^2
\cr
&\phantom{\biggl(}
-\Fr{1}{4 R^2}\left([X_\m,X_\n] +R^2[B_{\m\r},B_\n{}^\r]\right)^2
-\Fr{1}{2}[B_{\m\n},X^\n]^2
+R^2 D_+ B^{\m\n} D_- B_{\m\n}
\cr
&\phantom{\biggl(}
+iR^2 \chi^{\m\n}_- D_+\chi_{\m\n -}
+iR^2 \chi^{\m\n}_+ D_-\chi_{\m\n +}
+2iR^2 [B_{\m\n},\chi^{\m\n}_{+}]\xi_-
+2iR^2 [B_{\m\n},\chi^{\m\n}_-]\xi_+
\cr
&\phantom{\biggl(}
+2iR^2[C,\chi^{\m\n}_+]\chi_{\m\n-}
-R^2[C,B^{\m\n}][C,B_{\m\n}]
-2iR^2 B^{\m\n}[\chi_{\m\r+},\chi_{\n-}^{\phantom{-}\r}]
\cr
&\phantom{\biggl(}
-2i B_{\m\n}[\p^\m_+,\p^\n_-]
+2i\chi^{\m\n}_-[X_\m,\p_{\n +}]
-2i\chi^{\m\n}_+[X_\m,\p_{\n -}]
\biggr).
}
}
The resulting action is $Q_\pm$ invariant after modifying
\mcc\ and \nnw\ using the replacement \reppl.

\subsec{The Free RNS string Limit}

As a TFT, the path integral
is localized to the fixed point locus of global supersymmetry 
generated by $Q_\pm$. From \mcb, we can read off one important
equations,  the fixed point equation
$Q_\pm \xi_\mp =0$;
\eqn\saa{
F_{+-}=0.
}
Thus the path integral is always localized to the moduli space of 
flat
$U(N)$ connections. This will significantly simplify
our analysis since the connection can be gauged
away. Consider a Wilson line for the flat connection
\eqn\sab{
U_\g = P \exp \int_{\s_0}^{\s_0 + 2\pi} A_\s d\s,
}
which can be non-trivial.
Associating a Wilson line $\g\rightarrow U_\g$ to a non-contractable
loop $\g$ defines a homomorphism $\pi_1(S^1)\rightarrow U(N)$,
since  the parallel transformation along $\g$ depends
only on the homotopy class of $\r$. Conversely 
a homomorphism (or representation) $\rho:\pi_1(S^1)\rightarrow U(N)$
determines a rank $N$ flat vector bundle $E$.
Thus the moduli space of flat connections can be identified
with the representation variety modulo isomorphisms.
Of course $\pi_1(S^1)= \msbm{Z}$, as paths are classified 
by their winding number.
A representation (a $U(N)$ connection) 
can be either irreducible or reducible. 
In the latter case the vector
bundle decomposes into irreducible factors,
\eqn\sac{
E = E_1\oplus \cdots \oplus E_k.
}
Of course such a decomposition is
parameterized by the partition $N=\sum \n_k$ of the rank 
of the gauge group.
Equivalently, non-trivial Wilson lines break the $U(N)$
symmetry ($U(N)$ is broken down to the subgroup that commutes
with $U_\g$).

The other important fixed point equations, 
$Q_\pm \chi^{\m\n}_\mp=0$,
lead to the flat directions
\eqn\fcccd{
[X_\m,X_\n] +R^2[B_{\m\r},B_\n{}^\r]=0,\qquad 
[B_{\m\n},X^\n]=0,\qquad
[C,X^\m]=[C,B^{\m\n}]=0.
}
We can also examine the equations for fermionic zero-modes
from the action functional. By standard arguments in
TFT, we see that those equations are nothing but
the linearization of the fixed point equations
and the Coulomb gauge conditions. Since our model
is a BTFT, we do not have net $U$-number violation
in the path integral measure. In the present context
the $U$-number symmetry is just a part of ``world-sheet''
Lorentz invariance. 

In the limit $R^2=0$ we get the desired equations $[X_\m,X_\n]=0$.
This corresponds to the free RNS string limit.
All the $R^2$ dependent terms can be thrown away
and the theory is localized to configurations of commuting
matrices. Our balanced equivariant
cohomology generators $Q_\pm$ can be identified
with the left and right ``world-sheet'' supersymmetry.

We can rewrite the action functional $S$ defined in \dfg\ as
a one-parameter family of BTFT's
\eqn\wdfg{
S(R) =- Q_+Q_-\left(\int d\s^+ d\s^-
\biggl(iB^{\m\n}[X_\m,X_\n] + 2\p^\m_+ \p_{\m-}\biggr)\right)
+ R^2 Q_+ Q_-\left(\int d\s^+ d\s^- \CV\right),
}
where
\eqn\wpppf{
\eqalign{
&\CV =
-\tr\biggl(
iB^{\m\n}[B_{\m\r},B_{\n}{}^\s]+
\chi^{\m\n}_+\chi_{\m\n-}+ \xi_-\xi_+
\biggr).
}
}
We can regard $S(0)$ as
the action functional of $N$ copies of the free RNS string,
given by
\eqn\ftrial{
\eqalign{
S(0)&=\int d\s^+ d\s^- \tr\biggl(
2D_+X^\m D_-X_\m
+2i\p^{\m}_-D_+\p_{\m-}
+2i\p^{\m}_+D_-\p_{\m+}
+4i[C,\p^\m_+]\p_{\m-}
\cr
&\phantom{\biggl(}
+4i[X^\m,\p_{\m+}]\xi_-
+4i[X^\m,\p_{\m-}]\xi_+
-2[C,X^\m][C,X_\m]
+2H^\m H_\m
+2H_\m[B^{\m\n},X_\n]
\cr
&\phantom{\biggl(}
-H^{\m\n}[X_\m,X_\n]
-2i B_{\m\n}[\p^\m_+,\p^\n_-]
+2i\chi^{\m\n}_-[X_\m,\p_{\n +}]
-2i\chi^{\m\n}_+[X_\m,\p_{\n -}]
\biggr).
}
}
Here the anti-symmetric tensor multiplets are treated
as purely auxiliary fields. The integration over
$H_{\m\n}$ gives the delta-function like constraints
\eqn\wxx{
[X_\m,X_\n]=0,
}
so that our string coordinates $\{X^\m\}$ commute.
The $\chi^{\m\n}_\pm$ integration give further
delta function constraints
\eqn\wxy{
[X_\m,\p_{\pm\n}]=0,
}
which are the linearizations of \wxx\
We can also treat $(C,\xi_\pm)$ in a similar way.
The $\xi_\pm$ integrations give another delta-function
gauge constraint
\eqn\wxz{
[X^\m,\p_{\m \pm}]=0.
}
Finally the $B_{\m\n}$ together with $C$ integrations
give the constraints
\eqn\xya{
[\p^\m_+,\p^\n_-] =0,\qquad [H^\m,X^\n]=0.
} 
The constraints \wxy\ and \wxz\ are the linearization
of \wxx\ and the Coulomb gauge conditions respectively.
The last condition \xya\ is just the consistency condition.

>From \wxx, we see that $U(N)$ symmetry is generically broken down
to $U(1)^N$. Furthermore, the fixed point equations $Q_\pm \p^\m_\pm
=-iD_\pm X^\m=0$ imply that the $U(N)$ connections
should be reducible to have non-trivial solutions.\foot{
Note that the $X^\m$ are adjoint scalars.}
We can conclude that the action $S(0)$ is the straightforward 
formulation of a gas of free RNS string. 
In this formulation the off-diagonal part of $X^\m$
plays almost no role, except giving rise to a one-loop
determinant from to the localization. 

We can gauge away our connection
provided that we allow modified 
periodicity conditions
\eqn\sad{
X^\m(\t,\s_0 + 2\pi) = U_\g X^\m(\t, \s_0) U_\g^{-1}.
}
We can diagonalize
$X^\m = U x^\m U^{-1}$. Then the above action of the
Wilson line can be identified with  conjugation
$h x^\m h^{-1}$
of the  Weyl group $h$ on the eigenvalues $x^\m$ of $X^\m$.
Equivalently twisted sectors are parameterized
by the moduli space of flat connections.

Now we can follow the general arguments of DVV to interprete
our model as second-quantized free-string theory \DVV. 
As far as the bosonic fields are concerned, their arguments 
essentialy apply also to our model. The fermionic fields are 
much more difficult to treat. Especially the GSO projection 
we now need to impose gives some difficulties.

\subsec{Monad String as a Deformation of the RNS String Gas}

Now we can regard $S(R)$ as a deformation of $S(0)$
parameterized by $R$. Naively, such a deformation does
not change the theory since it is a pure $Q_\pm$ commutator.
However, the theory with $S(R)$ is only independent
of $R$, as clarified by Witten \WittenG, 
if (i) $S(R)$ has a non-degenerate kinetic energy
for all $R$; (ii) if there are no new fixed point to flow
in from infinity. Our choice of $\CV$ clearly does
not satisfy the above criterium. Turning on
$R$ introduce the kinetic terms for the anti-symmetric
tensor multiplets (via $\tr(\chi^{\m\n}_+\chi^{\m\n})$)
and the gauge multiplets (via $\tr(\xi_+\xi_-)$), as well 
as extra potential terms and Yukawa couplings for $B_{\m\n}$ 
(via $\tr(B^{\m\n}[B_{\m\r},B_\n{}^\r])$).
Furthermore it changes the fixed points \wxx\ via the
cubic term in $B^{\m\n}$.  Then
the off-diagonal parts of $X^\m$ will start to play
an important role due to the cubic term.

The above discussions also indicate that our construction
of the monad string is very natural, once we choose to
generalize the string coordinates to matrices.
We will see that it also directing us a more general theory
with $11$-dimensional covariance. 
It is also more natural to regard the theory
with $S(0)$ as a special limit of more fundamental
as general theory with $S(R)$. Thus we can interpret
the (ten-dimensional) monad string theory as an one-parameter 
family of theories, which reduces to the RNS string
in a special limit.

Remark that the terms in the action arising from $\CV$ 
in \wpppf\ lead to a well defined theory already by 
themselves, but only for the fields from the tensor 
and gauge multiplets. Really a similar action will be 
the starting point in the next section. 

\subsec{A Brief Comparison with the Matrix String Theory}

At this point, it will be usefull to compare with
matrix string theory. For example we can regard
$S(0)$ as the covariant RNS version of the free string
limit of matrix string theory. In matrix string theory,
as beautifully demonstrated by DVV \DVV,
the inverse of the two-dimensional Yang-Mill coupling
constant plays the role of type IIA string coupling 
constant.\foot{Note that matrix string theory is
two-dimensional $\CN=8$ physical super-Yang-Mills
theory. One the other hand, monad string is
a TFT in two-dimension and the Yang-Mill coupling
play no role. Note also that monad string is
not a twisted version of matrix string.}
In the monad string a similar role is given by $R^2$. 
However, there are some differences.

i) We will see that turning on $R^2$  directly
leads to a theory with $11$-dimensional covariance.
In the matrix string theory the relation with $11$-dimensional
theory is less direct.

ii) Turning on $R^2$ implies that 
the free monad strings start to couple with
the anti-symmetric tensor multiplets, which
are dynamical. In matrix string theory only the off-diagonal
parts of $X^\m$ are new contributions.

Considering the fact that matrix string theory is defined 
in the light-cone gauge, it is certainly possible
that monad string theory in the light-cone gauge  
is equivalent the matrix string theory.\foot{
This point was suggested to us by H.~Verlinde.}
We should also be very careful about
the role of anti-symmetric tensor
multiplets, which is absent in matrix string theory. 
Note that the Yukawa and potential terms are closely
related. We introduced the anti-symmetric
tensor multiplet to have the necesssary potential term while
maintaining $10$-dimensional covariance. 
In the matrix string (and in the light-cone GS formalism)
the counterparts of $\p^\m_\pm$ transform as
space-time spinors. Thus the covariant (at least 
in the light-cone gauge) form of Yukawa coupling
can be easily written down with the help of the soldering
form $\g^i_{a\dot{a}}$. The appearance of those crucial
central charges in the superalgebra is also due to the space-time
gamma matrices \BSb.
Clearly the anti-symmetric
tensor multiplet plays a similar role in our model.
Thus, it seems to be reasonable to believe that the anti-symmetric
tensor multiplet is the cost for a world-sheet supersymmetric
formulation and $10$-dimensional covariance.
On the other hand,
we will see that the anti-symmetric tensor multiplet
is very important and more fundamental in the $11$-dimensional
viewpoints.
 
In the next subsection, we will briefly examine a possible
interpretation of the anti-symmetric tensor multiplet from
in the ten-dimensional view-point.

\subsec{Another Perturbation}

As shown earlier, the anti-symmetric tensor
multiplet can be regarded as purely auxiliary fields
as long as we set $R=0$. However, we have seemingly
mysterious equations 
\eqn\mise{
[X_\m, B^{\m\n}]=0,
}
from \ftrial\ even in the free string limit.
We also note that free monad string theory
can not be defined without the $B^{\m\n}$-multiplet.
However, we can define a free theory of $B^{\m\n}$
without the help of the $X^\m$-multiplet. The action
functional can be defined
as
\eqn\misa{
S_B = -Q_+Q_-\int d\s^+ d\s^- \CV,
}
where $\CV$ is given by \wpppf. 
This can be regarded as a clue that something
described by the anti-symmetric
tensor multiplet is more fundamental 
than string itself.

Now we will consider yet another deformation.
We consider
\eqn\maab{
S(R,m) = S(R) - m Q_+Q_- \int d\s^+ d\s^-\tr\left(
\Fr{i}{2}B^{\m\n}B_{\m\n}\right),
}
where $m$ plays the role of a bare mass
for the anti-symmetric tensor multiplet.
We have
\eqn\mnn{
\eqalign{
S(R,m) =& \int\tr\left(
R^2 H^{\m\n}H_{\m\n} 
-H^{\m\n}\left([X_\m,X_\n]+R^2[B_{\m\r},B_{\n}{}^\r] 
-m B_{\m\n}\right)
+i m \chi^{\m\n}_{+}\chi_{\m\n-}\right)
\cr
&+\ldots.
}
}
We can eliminate $H^{\m\n}$ by setting
\eqn\retty{
H_{\m\n} = \Fr{1}{2}\biggl([X_\m, X_\n]+R^2[B_{\m\r},B_{\n}{}^\r] 
-m B_{\m\n}\biggr),
}
we get
\eqn\mnnd{
\eqalign{
S^\pr(R,m) = & S^\pr(R)+ \int d^2\s\tr\left(
\Fr{m}{2}B^{\m\n}
 \left([X_\m, X_\n]+R^2[B_{\m\r},B_{\n}{}^\r]\right) 
+im\chi^{\m\n}_{+}\chi_{\m\n-}\right)
\cr
&
-\Fr{m^2}{4} \int d^2\s\tr\left(B^{\m\n}B_{\m\n}\right),
}
}
where $S^\pr(R)$ is given by \ttrial.
It is also understood that we integrated out $H^\m$ by
setting
\eqn\vcd{
H_\m = -[B_{\m\n},X^\n].
}

This simple looking perturbation is very interesting.
The theory is localized to the flat directions
given by $H_{\m\n}=H^\m=0$;
\eqn\fftr{
\eqalign{
[X_\m, X_\n]+R^2[B_{\m\r},B_{\n}{}^\r] 
-m B_{\m\n}=0,\cr
[X^\m, B_{\m\n}]=0,\cr
}
}
Combining these equations, we also have
\eqn\flatd{
\Fr{1}{m}[X^\m,[X_\m, X_\n]] 
+\Fr{R^2}{m}[X^\m,[B_{\m\r},B_{\n}{}^\r]]=0.
}
Now we will consider two examples.

i) Consider a particular sector of our moduli space
such that $\{X^\m\}$ commutes with each others.
We have
\eqn\poincare{
\eqalign{
[X_\m, X_\n]&=0,\cr
[X^\m, B_{\m\n}]&=0,\cr
[B_{\m\r},B_{\n}{}^\r] &=\Fr{m}{R^2}B_{\m\n}.\cr
}
}

ii) For $R=0$ we can eliminate $B^{\m\n}$ 
from $S(0,m)$ by setting
\eqn\vbb{
B_{\m\n} =\Fr{1}{m}[X_\m,X_\n].
}
Then the flat direction is given by
\eqn\vfcv{
[X^\m, [X_{\m},X_\n]]=0.
}
We can also eliminate $\chi^{\m\n}_\pm$ by the simple
algebraic equation of motion.
Then the action functional $S(0,m)$ can be
written as
\eqn\mbv{
S(0,m)= -\Fr{1}{2}(Q_+Q_- -Q_-Q_+)\left(\int d\s^+ d\s^-
\tr\left(-\Fr{1}{2m}[X^\m,X_\n][X_\m,X_\n] +\p^\m_+\p_{\m-}
\right)\right).
}
where 
the replacement of $H^\m$ in \mcc\ with
\vcd\ is understood.
For finite $N$ $S(0,m)$ is equivalent to the unperturbed
theory $S(0,0)$.
For $N\rightarrow \infty$ and if we want to
change the commutators to Poisson brackets 
we will have higher critical points in \vfcv.

With an analogy to the matrix
theory, the relation \vbb\ seems to suggest that $B_{\m\n}$
is somehow related to membrane.

\newsec{Eleven Dimensional Covariance}

In this section we will provide a more fundamental
description. The starting point is an observation that
the two multiplets $(X^\m,\p^\m_\pm,H^\m)$ and 
$(B^{\m\n}, \chi^{\m\n}_\pm, H^{\m\n})$ can be
naturally combined into a single multiplet, which
transform as an anti-symmetric second rank tensor under
$SO(10,1)$. We will suggest that the resulting theory
is a formulation of the sought for M theory.

\subsec{The Algebra}

Let $\S$ be a $(1+1)$-dimensional 
cylinder $S^1\times R$ with light-cone coordinates
$\s^\pm =\Fr{1}{2}(\s \pm \t)$.
Let $I,J,K,L=0,1,\ldots,10$. We introduce
an adjoint ``world-sheet'' scalar field $B^{IJ}(\s^+,\s^-)$, which 
transforms as an anti-symmetric second rank tensor under
$SO(10,1)$. 
We denote by $g^{IJ}$ the usual Minkowski metric
in $\msbm{R}^{10,1}$.
The $U(N)$ local gauge symmetry acts on $B^{IJ}$ as
\eqn\zab{
B^{IJ}\rightarrow g B^{IJ} g^{-1},\qquad g:U(N)\rightarrow \S.
}
In the space of all fields $B^{IJ}$ we introduce a
natural gauge invariant and $SO(10,1)$ invariant metric
\eqn\zaa{
\left|\d B\right|^2 =\int d\s^+ d\s^-\tr\left(B^{IJ}B_{IJ}\right).
}
Although the direct geometrical interpretation is obscure, we will
still refer to $B^{IJ}$ as the ``space-time'' coordinates
of ``strings'' in eleven-dimensions.
With the above basic setting, we will construct the unique theory
in an unbroken phase of the $11$-dimensional covariance.
Note that we are not imposing the ``space-time'' super-Poincar\'e
symmetry.
The $\CG\times P_{\s^\pm}$-equivariant cohomology algebra
is given by
\eqn\ennw{
\eqalign{
&Q_+ B^{IJ} = i\chi^{IJ}_+,\cr
&Q_- B^{IJ} = i\chi^{IJ}_-,\cr
}\qquad
\eqalign{
&Q_+ \chi^{IJ}_+ = -D_+ B^{IJ},\cr
&Q_+ \chi^{IJ}_- = + H^{IJ} - [C,B^{IJ}],\cr
&Q_- \chi^{IJ}_+ = - H^{IJ} - [C,B^{IJ}],\cr
&Q_- \chi^{IJ}_- = -D_-B^{IJ},\cr
}
}
satisfying the commutation relations 
\eqn\vaa{
Q_\pm^2 B^{IJ}= -i \rd_\pm B^{IJ} -i[\phi_{\pm\pm}, B^{IJ}],
\qquad \{Q_+,Q_-\}B^{IJ} =-i[C,B^{IJ}].
}
We can interpret $i\chi^{IJ}_\pm$ as ghosts associated
with the symmetry under arbitrary shift $B^{IJ}\rightarrow
B^{IJ} +\d B^{IJ}$. As usual $Q_\pm^2=0$ modulo the gauge
transformation generated by $\phi_{\pm\pm}$ as well
as the ``world-sheet'' translation along 
the $\s^\pm$ directions, i.e.,
modulo the redundancy of our system. $\{Q_+, Q_-\}=0$
modulo gauge transformation generated by $\phi_{+-}=C$.
As earlier $\phi_{\pm}$ are the left and right components
of an $U(N)$ connection and $C$ is an
adjoint scalar on the ``world-sheet'' with the $Q_\pm$
algebra given by \mcb.
The auxiliary fields
$H^{IJ}$ transform as
\eqn\ebbg{\eqalign{
&Q_+ H^{IJ}= -i D_+\chi^{IJ}_- +i[C,\chi^{IJ}_+] 
             +i[\xi_+, B^{IJ}],\cr
&Q_- H^{IJ}= +i D_-\chi^{IJ}_+ -i[C,\chi^{IJ}_-] 
             -i[\xi_-, B^{IJ}].\cr
}
}
We have $sl_2$ symmetry and an associated additive quantum 
number $U$
of the above algebra, which can be summarized as usual
\eqn\mymvi{
\matrix{
U=+1\cr
{}\cr
U=0\cr
{}\cr
U=-1\cr
{}\cr
}\qquad
{\rm fields}\quad
\matrix{    &  &   \chi^{IJ}_{+}  &  &  \cr
    & \nearrow &  &  \searrow & \cr
B^{IJ} &  & & & H^{IJ}\cr
 & \searrow & & \nearrow & \cr
 & & \chi^{IJ}_{-}& & \cr},
}
We will call the above multiplet the $11$-dimensional 
anti-symmetric tensor multiplet.

\subsec{The Action Functional}

Now we define an almost unique $SO(10,1)$ and $sl_2$ as well
as gauge invariant
action functional by
\eqn\elaction{
S_{11} =-Q_+ Q_- \int d^+\s d\s^-\tr\biggl(
\Fr{i}{3}B^{IJ}[B_{IK},B_J{}^{K}]
+\chi^{IJ}_+\chi_{IJ-} +\xi_-\xi_+\biggr).
}
The global $SO(10,1)$ and $sl_2$ symmetries may be interpreted
as the ``space-time'' and ``world-sheet'' Lorentz symmetries,
respectively. One can regard the above action functional
as a BRST quantized version of an underlying theory with
unbroken $11$-dimensional general covariance.
We have
\eqn\elex{
\eqalign{
S_{11}=&\
\int d\s^+ d\s^- \tr\biggl(
D_+ B^{IJ} D_- B^{IJ}
+i\chi^{IJ}_- D_+\chi_{IJ -}
+i\chi^{IJ}_+ D_-\chi_{IJ +}
+2i[C,\chi^{IJ}_+]\chi_{IJ-}
\cr
&\phantom{\biggl(}
+{2i} [B_{IJ},\chi^{IJ}_{+}]\xi_-
+{2i} [B_{IJ},\chi^{IJ}_-]\xi_+
-[C,B^{IJ}][C,B_{IJ}]
+H^{IJ}H_{IJ}
\cr
&\phantom{\biggl(}
-H^{IJ}[B_{IK},B_J{}^{K}]
-{2i}B^{IJ}[\chi_{IK+},\chi_{J-}^{\phantom{-}K}]
-D_+ C D_- C
\cr
&\phantom{\biggl(}
-i\xi_-D_+\xi_-
-i\xi_+D_-\xi_+
+2i\xi_+[C,\xi_-]
-\Fr{1}{4}F_{+-}^2
\biggr).
}
}
Integrating out $H^{IJ}$ by setting
\eqn\rrpl{
H_{IJ}=\Fr{1}{2}[B_{IK},B_J{}^K],
}
we have
\eqn\elext{
\eqalign{
S_{11}=&
\int d\s^+ d\s^- \tr\biggl(
D_+ B^{IJ} D_- B^{IJ}
+i\chi^{IJ}_- D_+\chi_{IJ -}
+i\chi^{IJ}_+ D_-\chi_{IJ +}
-D_+ C D_- C
\cr
&
-i\xi_-D_+\xi_-
-i\xi_+D_-\xi_+
+2i[B_{IJ},\chi^{IJ}_{+}]\xi_-
+2i[B_{IJ},\chi^{IJ}_-]\xi_+
+2i[C,\chi^{IJ}_+]\chi_{IJ-}
\cr
&
+2i\xi_+[C,\xi_-]
-2iB^{IJ}[\chi_{IK+},\chi_{J-}{}^{K}]
-\Fr{1}{4}[B^{IK}, B_{JK}][B_{IL},B^{JL} ]
\cr
&
-[C,B^{IJ}][C,B_{IJ}]
-\Fr{1}{4}F_{+-}^2
\biggr).
}
}
Now the transformation law \ennw\ should be changed to
\eqn\qqa{
\eqalign{
&Q_+\chi^{IJ}_- =+\Fr{1}{2}[B^{IK},B^J{}_K]-[C,B^{IJ}],
\cr
&Q_-\chi^{IJ}_+ =-\Fr{1}{2}[B^{IK},B^J{}_K]-[C,B^{IJ}],\cr
}
}
The above modification preserves our commutation relations, i.e.,
$Q^2_\pm\chi^{IJ}_\mp= -iD_\pm\chi^{IJ}_\mp$,
provided that
\eqn\eqmc{
\eqalign{
iD_-\chi^{IJ}_+ 
-    i     [B^{IK},\chi_{K-}^{\phantom{-}J}]
-    i     [C,\chi^{IJ}_-]
-    i     [\xi_-,B^{IJ}]
=0,\cr
iD_+\chi^{IJ}_- 
+    i     [B^{IK},\chi_{K+}^{\phantom{-}J}]
-    i     [C,\chi^{IJ}_+]
-    i     [\xi_+,B^{IJ}]
=0,\cr
}
}
which are just the equations of motion of $\chi^{IJ}_\pm$.
The $B^{IJ}$ equation of motion is
\eqn\qqb{
\Fr{i}{2}D_+ D_-B^{IJ}-\Fr{i}{2}[B^{IK},[B_{KL},B^{JL}]]
+[\chi^{IK}_+,\chi_{J-}^{\phantom{-}K}]
+[\xi_+,\chi^{IJ}_-]
+[\xi_-,\chi^{IJ}_+]
=0,
}
which is a supersymmetry variation of \eqmc.

\subsec{Back to  Ten Dimensions}

Now we will break the eleven-dimensional covariance
down to the ten-dimensional one.
{}From now on we will label the $SO(10,1)$ vector indeces
$I,J,K,L =1,\ldots,11$.
We fix
the $11$-dimensional metric 
$g^{IJ} = \left(\matrix{g^{\m\n}&0\cr 0& g^{11\,11}}\right)$
with $g^{11\,11}= 1/R^2$.
Then we define a $10$-dimensional (non-commutative) 
coordinate by $X^\m=B^{11\m}$. 
Similarly, we set $\p^\m_\pm=\chi^{11\m}_\pm$ and
$H^\m = H^{11\m}$. The supersymmetry algebras \mca, \mcc, \mcd, 
\nnw,
and \nnww\ follow from \ennw\ and \ebbg.
Then the action \elaction\ reduces to
\eqn\tenact{
\eqalign{
S_R=-Q_+ Q_- \int d^2\s &\tr\biggl(
i B^{\m\n}\left(
[X_\m,X_\n]
+\Fr{R^2}{3}[B_{\m\r},B_\n{}^{\r}] 
\right)
\cr
&
+R^2\chi^{\m\n}_+\chi_{\m\n-} + 2\p^\m_+\p_{\m -} 
+R^2 \xi_-\xi_+\biggr),
}
}
where we scaled the action by an overall factor $R^2$.
The above action is exactly the same as \dfg, for which the 
explicit form is given by \trial\ and \ttrial.
As discussed in Sect.~$4.3$ and $4.4$, perturbation away
from free strings directly lead us the eleven
dimensional picture.

At first sight the free string limit looks  counter intuitive.
It corresponds to the infinite radius $1/R$ of $11$th direction of
the background $11$-dimensional space. Furthermore, it is
natural to identify the string coupling constant with
$R^2$. Since however we do not have $11$-dimensional 
string coordinates, 
the above problem should be reexamined. 

First we need to clarify our usage of
``compactification'' to a circle. For practical
purpose the equation \tenact\ is the definition
of compactification on a circle of our eleven dimensional
theory \elaction. 
In any ``world-sheet'' formulation of superstring theory,
the space-time Lorentz invariance is detected
by global symmetry. In terms of $SO(9,1)$
acting on indices $\m,\n=1,\ldots,10$, $B^{11\m}$
transform as a vector and $B^{\m\n}$ transforms
as an anti-symmetric tensor. Since we do not
have the $11$-th component of the vector (or string coordinates),
we can not impose any other conditions apart from the
form of the background metric.
As for $10$-dimensional vectors $B^{11\m}(\s,\t)$
we may use those as certain ``string coordinates''
in ``space-time''. From our viewpoint, any space-time
interpretation is just an effective description.
The most reasonable description of the model defined
by \tenact\ is to regard it as a family of theories
parameterized by $R^2$.

Note that we have manifest $11$-dimensional covariance.
However, we do not have the usual coordinate interpretation
of $11$-dimensional space-time. Only after the
reduction to $10$ dimensions we get (non-commutative)
coordinates of strings. Now the most difficult question
is if our model has a particular limit where an $11$-dimensional
space-time picture appears. Provided that our model
describes M theory, we should certainly expect
this \WittenV. Finding the free string theory as an effective
description is very easy in our approach. However,
the appearance of $11$-dimensional supergravity
can be a very difficult quantum mechanical 
problem. At this point, we will just leave the
difficult problems for the future.

Our approach has another difficult problem.
Up to now we did not worry
about the GSO projection. It is of no doubt that we need
the GSO projection in the free string limit. 
We expect that the quantum mechanical consistency
of our model may determine a particular projection
for a particular free string limit.
We do not know any direct justification for the above
wishful thinking. In the next subsection, we will study 
our model after compactifying further down
to a circle. We will show that our model has
two types of string limits, which behave
as type IIA and type IIB strings, as well as the predictions
based on M theory viewpoints \AS. 
We may use the examples
as the evidence for that our model after proper
quantization automatically decide a particular
GSO projection at a particular limits.

\subsec{A Further Compactification}

We can compactify our model further. We will now study the model 
when compactified on a $T^2$ in the $10-11$ direction.
The background metric is given by 
\eqn\bmnb{
g^{IJ}=\left(\matrix{
g^{ij} & 0 &0\cr
0 &\Fr{1}{R_{10}^2}&0\cr
0& 0& \Fr{1}{R_{11}^2}}
\right)
} 
The index $i$ will refer to the first 9 uncompactified directions. 
Then we have $2$ ``space-time coordinates'' 
of strings instead of the one $X^i$ from the 
last section. These we denote
\eqn\qwa{
X^i_{(11)} = B^{11i},
\qquad
X^i_{(10)} = B^{10i}.
}
They will have superpartners $\p_{(1)\pm}^i$ and 
$\p_{(2)\pm}^i$ respectively. Furthermore there is 
a 9 dimensional scalar $\phi=B^{10\, 11}$ 
with superpartners $\th_\pm$. We can summarize the 
supersymmetry
by the following diagram
\eqn\nnb{
\matrix{    &  &   \p^{i}_{(a)+}  &  &  \cr
    & \nearrow &  &  \searrow & \cr
X^{i}_{(a)} &  & & & H^{i}_{(a)}\cr
 & \searrow & & \nearrow & \cr
 & & \p^i_{{(a)}-}& & \cr},
\qquad 
\matrix{    &  &   \chi^{ij}_{+}  &  &  \cr
    & \nearrow &  &  \searrow & \cr
B^{ij} &  & & & H^{ij}\cr
 & \searrow & & \nearrow & \cr
 & & \chi^{ij}_{-}& & \cr},
\qquad
\matrix{    &  &   \th_{+}  &  &  \cr
    & \nearrow &  &  \searrow & \cr
\phi &  & & & H \cr
 & \searrow & & \nearrow & \cr
 & & \th_{-}& & \cr}.
}
Note that we can combine the $\phi$-multiplet either with
the $X^i_{(11)}$ or with $X^i_{(10)}$ multiplet to get 
ten-dimensional
multiplets $X^\m_{(11)}$ and $X^\m_{(10)}$, 
respectively, at the decompactification
limit of one of the directions.

The action will then depend on the parameters $R_{10}$ and
$R_{11}$ as
\eqn\nineact{
\eqalign{
S_9=- &Q_+ Q_-\!\! \int d^2\s \tr\biggl(
2R_{10}^2\p^i_{(11)+}\p^{(11)}_{i -}
+ 2R_{11}^2\p^i_{(10)+}\p^{(10)}_{i -}
+(R_{10}R_{11})^2\chi^{ij}_+\chi_{ij-} 
\cr
&
+iB_{ij}\left(
R_{10}^2[X^i_{(11)},X^j_{(11)}]
+R_{11}^2[X^i_{(10)},X^j_{(10)}]
+\Fr{(R_{10}R_{11})^2}{3}[B_{ik},B_j{}^{k}] 
\right)
\cr
&
+4i\phi[X^i_{(11)},X_i^{(10)}] +\th_+\th_-
+ (R_{10}R_{11})^2\xi_-\xi_+
\biggr).
}
}
This action has one obvious symmetry by exchanging
the first and the second
$9$-dimensional vector multiplets
accompanied with
$R_{10}\leftrightarrow R_{11}$. This symmetry
came from the underlying eleven-dimensional covariance
of our model. 

We can regard the action $S(R_{10},R_{11})$  
as describing the two-parameter family 
of theories. 
To explore the moduli space we consider the potential term
\eqn\qwb{
V_{11} = \Fr{1}{4}\tr\biggl(
[B^{IK}, B_{JK}][B_{IL},B^{JL} ]
\biggr),
}
and the flat directions $V_{11}=0$.
In nine dimensions we have
\eqn\qwc{
\eqalign{
V_9=&\tr\biggl(
\Fr{1}{4}
\left(
\Fr{R_{11}}{R_{10}}[X_i^{(10)},X_j^{(10)}]
+\Fr{R_{10}}{R_{11}}[X_i^{(11)},X_j^{(11)}]
+R_{10}R_{11}[B^{ik}, B_{jk}]
\right)^2
\cr
&
+\Fr{1}{2}\left(
 \Fr{1}{R_{11}}[\phi,X_i^{(11)}] 
 -R_{11} [B_{ik}, X^k_{(10)}]\right)^2
+\Fr{1}{2}\left(
 \Fr{1}{R_{10}}[\phi,X_i^{(10)}] 
 +R_{10} [B_{ik}, X^k_{(11)}]\right)^2
\cr
&
+\Fr{1}{2}[X^i_{(11)},X_i^{(10)}]^2
\biggr),
}
}
leading to the following flat directions
\eqn\qwd{
\eqalign{
\Fr{R_{11}}{R_{10}}[X_i^{(10)},X_j^{(10)}]
 +\Fr{R_{10}}{R_{11}}[X_i^{(11)},X_j^{(11)}]
 +R_{10}R_{11}[B^{ik}, B_{jk}]=0,\cr
\Fr{1}{R_{11}}[\phi,X_i^{(11)}] 
 -R_{11} [B_{ik}, X^k_{(10)}]=0,\cr
\Fr{1}{R_{10}}[\phi,X_i^{(10)}] 
 +R_{10} [B_{ik}, X^k_{(11)}]=0,\cr
[X^i_{(11)},X_i^{(10)}]=0.
}
}

Now we examine special points in our moduli
space where the usual string pictures appears.

1) We consider the limit $R_{11}=0$ and
$R_{10}\rightarrow \infty$. This reduces to the free
string limit discussed in the previous subsection.
{}From the second equation in \qwd, we
see that $\phi$ commutes with $X^i_{(11)}$
to form  the string coordinates represented by $X^\m_{(11)}$.
Similarly in the limit $R_{11}=\infty$ and $R_{10}=0$,
we get another ten-dimensional strings with coordinates
$X^\m_{(10)}$. In both cases the $B_{ij}$-multiplets
are completely decoupled. 
Those should correspond to the limits for two equivalent 
IIA strings.

2) Now we consider limit that $R_{10},R_{11}\rightarrow 0$
while taking $R_{10}/R_{11}$ arbitrary. 
Then our crucial equations \qwd\ reduce
to 
\eqn\qwdb{
\eqalign{
\Fr{R_{11}}{R_{10}}[X_i^{(10)},X_j^{(10)}]
 +\Fr{R_{10}}{R_{11}}[X_i^{(11)},X_j^{(11)}]
 =0,\cr
[\phi,X_i^{(11)}] =0,\cr
[\phi,X_i^{(10)}] =0,\cr
[X^i_{(11)},X_i^{(10)}]=0.
}
}
Now $\phi$ commutes with both $X^{(11)}_i$ and $X^{(10)}_i$.
 But it is in
either the $R_{10}/R_{11} \rightarrow 0$
or the $R_{10}/R_{11} \rightarrow 0$ limit that
one of these coordinates describes free strings.
The action functional is effectively given by
\eqn\nineact{
\eqalign{
S_9=&- Q_+ Q_- \int d^2\s \tr\biggl(
2R_{11}^2\p^i_{(10)+}\p^{(10)}_{i -} 
+ 2R_{10}^2\p^i_{(11)+}\p^{(11)}_{i -} 
+\th_+\th_-
\biggr)\cr
&+\Fr{1}{4}\int d^2\s 
\tr\left(
\Fr{R_{10}}{R_{11}}[X_i^{(10)},X_j^{(10)}]
+\Fr{R_{11}}{R_{10}}[X_i^{(11)},X_j^{(11)}]
\right)^2 
\cr
&+\Fr{1}{2}\int d^2\s  
\tr\left(
\Fr{1}{R_{11}^2}[\phi,X^{(11)}_i]^2
+\Fr{1}{R_{10}^2}[\phi,X^{(10)}_i]^2
\right)
\cr
&
+Yukawa
}
}
where the Yukawa term has a pattern similar to
 the potential term.
It is also clear that one system is strongly
coupled if the other is weakly coupled.
Because of the obvious symmetry 
$X_i^{(10)}\leftrightarrow X_i^{(11)}$
and $R_{10}\leftrightarrow R_{11}$, we have manifestly
self-dual system. Whatever system we are describing
we find one with manifest and non-perturbative
$S$-duality. 
These limits of our model should  correspond to the type IIB
strings. Perturbatively we will only see the usual 
string action, arising from only one of the sets of coordinates.
 But in general we find contributions from both of them, and the
usual space-time interpretation breaks down. 

It will be interesting to see how our approach
can be generalized so to give rise to the heterotic
and type I strings \HW. This may be done using 
procedures similar to those in matrix string theory \Mhet.

\subsec{A Further Generalization}

As argued earlier, it seems that $B^{IJ}(\s,\t)$ is 
related to the membrane of M theory. 
Here we merely refer to the membrane M theory 
as certain degrees of freedom which are required
to produce the string theoretic degrees of freedom
in lower dimensions after double compactification. 
>From the viewpoint
of an observer living in the lower dimensions, certain components
of $B^{IJ}$ behave as ``space-times coordinates''
of strings. 

We also expect to have the five-branes of M theory
in $11$-dimensions.
Following the previous discussions, we
mean by a five-brane in eleven dimensions an object
which transforms as an anti-symmetric $5$ rank tensor
under $SO(10,1)$. After breaking the $11$-dimensional
covariance down to the $7$-dimensional one, for example, 
it can be identified with the ``space-time'' coordinates
of strings. We introduce a rank $5$ anti-symmetric
tensor $J^{IJKLM}(\s,\t)$ which are ``world-sheet''
adjoint scalars. We have the usual supermultiplet
\eqn\qwl{
\matrix{    &  &   \Bv^{IJKLM}_{+}  &  &  \cr
    & \nearrow &  &  \searrow & \cr
J^{IJKLM} &  & & & H^{IJKLM}\cr
 & \searrow & & \nearrow & \cr
 & & \Bv^{IJKLM}_{-}& & \cr},
}
and the corresponding super-algebra.

Now we are looking for a cubic action potential
term to write down the potential term for $J^{IJKLM}$.
There is no $SO(10,1)$ invariant cubic terms for $J^{IJKLM}$.
The only possibility is to couple with the
cubic action potential of $B^{IJ}$. So we have a
more or less unique choice as usual, given by
\eqn\elaction{
\eqalign{
S_{11}(\b) =-Q_+ Q_- &\int d\s^+ d\s^-\tr\biggl(
\Fr{i}{3}B^{IJ}\biggl([B_{IK},B_J{}^{K}] +3{\b}
[J_{IKLMN},J_{J}{}^{KLMN}]\biggr)\cr
&
+\chi^{IJ}_+\chi_{IJ-} 
+\b \Bv^{IJKLM}_+\Bv_{IJKLM-}
+\xi_-\xi_+
\biggr),
}
}
where $\b$ is a new coupling constant.

If we write down the action explicitely, we have
\eqn\elex{
\eqalign{
S_{11}(\b)&=
\int d\s^+ d\s^- \tr\biggl(
D_+ B^{IJ} D_- B_{IJ}
+i\chi^{IJ}_- D_+\chi_{IJ -}
+i\chi^{IJ}_+ D_-\chi_{IJ +}
+2i[C,\chi^{IJ}_+]\chi_{IJ-}
\cr
&\phantom{\biggl(}
+{2i} [B_{IJ},\chi^{IJ}_{+}]\xi_-
+{2i} [B_{IJ},\chi^{IJ}_-]\xi_+
-[C,B^{IJ}]^2
+\b D_+ J^{IJKLM} D_- J_{IJKLM}
\cr
&\phantom{\biggl(}
+i\b \Bv^{IJKLM}_- D_+\Bv_{IJKLM -}
+i\b\Bv^{IJ}_+ D_-\Bv_{IJKLM +}
+2i\b[C,\Bv^{IJKLM}_+]\Bv_{IJKLM-}
\cr
&\phantom{\biggl(}
+{2i}\b [J_{IJKLM},\Bv^{IJKLM}_{+}]\xi_-
+{2i}\b [J_{IJKLM},\Bv^{IJKLM}_-]\xi_+
-\b[C,J^{IJKLM}]^2
\cr
&\phantom{\biggl(}
+H^{IJ}H_{IJ}
-H^{IJ}\left([B_{IK},B_J{}^{K}] 
+ \b[J_{IKLMN},J_{J}{}^{KLMN}]\right)
+\b H_{IJKLM}^2
\cr
&\phantom{\biggl(}
+2\b H_{IKLMN}[B^{IJ}, J_{J}{}^{KLMN}]
-2i\b B^{IJ}[\Bv_{IKLMN+},J_{J-}^{\phantom{-}KLMN}]
\cr
&\phantom{\biggl(}
+2i\b\chi^{IJ}_-[\Bv_{IKLMN+},J_{J}{}^{KLMN}]
-2i\b\chi^{IJ}_+[\Bv_{IKLMN-},J_{J}{}^{KLMN}]
\cr
&\phantom{\biggl(}
-{2i}B^{IJ}[\chi_{IK+},\chi_{J-}^{\phantom{-}K}]
-D_+ C D_- C
-i\xi_-D_+\xi_-
-i\xi_+D_-\xi_+
\cr
&\phantom{\biggl(}
+2i\xi_+[C,\xi_-]
-\Fr{1}{4}F_{+-}^2
\biggr).
}
}
We integrate out $H^{IJ}$ and $H^{IJKLM}$ by setting
\eqn\eerrpl{
\eqalign{
&H_{IJ}=\Fr{1}{2}[B_{IK},B_J{}^K] +\Fr{\b}{2}
[J_{IKLMN},J_{J}{}^{KLMN}],\cr
&H^{IKLMN}=-[B^{IJ}, J_{J}{}^{KLMN}]
}
}

\subsec{Universal Monads and M Theory}

Now the two equations
\eqn\umonads{
\eqalign{
[B_{IK},B_J{}^K] +\b
[J_{IKLMN},J_{J}{}^{KLMN}]=0,\cr
[B^{IJ}, J_{J}{}^{KLMN}]=0,\cr
}
}
which define the flat directions, are the most
important equations we have. We will call
a set of matrices $(B,J)$ satifying \umonads\
a {\it universal monad}. For a constant
monad, we may associate a {\it universal instantons}.
The equation \umonads\ is the end point
of our generalization of the simple matrix equations
$[X^\m,X^\n]=0$ describing point-like instantons
in $10$-dimensions.

Our model is classified by the space of
solutions, modulo gauge equivalence, of
\umonads.  Our conjecture
that we are describing $M$-theory
means that the moduli
space is identical to that
of  M theory. According to our conjecture
all the information of  strings and other
extended objects should be encoded
in \umonads. Furthermore, as the moduli
space of theories, we should be able to
find special points where the known string
theories are the effective descriptions.
We also expect the web of string dualities
to be manifest as the symmetry
in the bulk.
After compactification to lower dimensions we will
get a much richer structure of the moduli space.
By examining the corresponding reduction
of the equation \umonads\ we should be able
to find numerous theories
and mutual relations with eachother.

The detailed examination of the
entire moduli space defined by \umonads\
is beyond the scope of this paper.
We merely want to point out that
the theory compactified on
$T^4$ should be very interesting. It is the first
dimension where some components of $J^{IJKLM}$
transform as $SO(6,1)$ vectors which give rise to
new a set the ``space-time coordinates'' of strings.
Compactifying further down to $T^5$,
we have $5$-sets of string coordinates
from $B^{IJ}$ and another $5$-sets of string
coordinates from $J^{IJKLM}$. 
By examining the corresponding reduction
of \umonads, which describe a $5$-dimensional
space of theories, we will be able to discover
the various different theories. The mutual
relations between those theories should
follow from a very easy analysis.
This may be related with new phenomena
in M theory compactified on $T^4$ and $T^5$ \BRS.

\newsec{Further Points to Examine}

There are several important issues we ignored in our analysis. 
First of all, what is the underlying 
geometrical interpretation of $B^{IJ}(\s,\t)$ 
and $J^{IJKLM}(\s,\t)$? We already mentioned a possible 
connection with the membrane and fivebrane of M theory. 
How this relation comes about we do not know at the moment.

We restricted our attention to classical considerations. 
The quantization surely will introduce some delicate issues:

1) 
The spacetime supersymmetry and GSO projection;
For the free string limit
we certainly have spacetime supersymmetry.
However it is not obvious that the spacetime 
supersymmetry is a generic property of our model 
in any situation. Our construction indicates that the 
spacetime interpretation itself is an effective description.
Even in the free string limit we need to
impose a GSO projection to obtain spacetime
supersymmerty. Since the free strings are
embbedded into a bigger picture in our model there should
be a generalized notion of GSO projection.
We speculated that a proper GSO
projection could arise via certain quantum
consistencies at a particular point in the
moduli space. If our speculation is correct,
the spacetime supersymmetry itself can be viewed
as an effective description.
These issues are closely related with
the notions of the unbroken and broken
phases of general covariance.
According to a purely classical argument,
our model should not contain gravitons.
However, we found special points corresponding
to free strings where gravitons certainly exist.
All these issues seem to be subtle and difficult
quantum mechnical properties.

2) 
The critical dimension;
It is only the $11$-dimensional model, 
as constructed in this paper, that gives 
rise to crtitical strings. 
Since the string appears as an effective 
description, the usual notion of critical dimension
can be meaningless. At least classically, 
our model can be formulated in arbitrary
dimensions with arbitray field contents.
Hopefully, the conditions for a consistent quantum theory 
lead us to the correct dimension for our model.

\ack{
It is our pleasure to thank Erik and Herman Verlinde
for illuminating discussions and comments
as well as for encouragements.
This work is supported by FOM 
 and a pioneer fund
of NWO.

\listrefs

\bye